\documentclass[11pt]{article}

\usepackage[utf8]{inputenc}
\usepackage{color}
\usepackage{mathtools}
\usepackage[font=footnotesize,labelfont=bf]{caption}
\usepackage{authblk}
\usepackage{blindtext}
\usepackage{hyperref}
\DeclareMathAlphabet{\mathpzc}{OT1}{pzc}{m}{it}
\expandafter\let\csname equation*\endcsname\relax
\expandafter\let\csname endequation*\endcsname\relax
\usepackage{amsmath}

\usepackage{color}
\usepackage{hyperref}
\makeatletter
\newcommand\xlabel[2][]{\phantomsection\def\@currentlabelname{#1}\label{#2}}
\makeatother
\usepackage{soul,xcolor}
\setstcolor{red}
\usepackage{amsmath}
\usepackage{amssymb}

\title{Modelling co-evolution of resource feedback and social network dynamics in human-environmental systems.
}

\author[1]{Meghdad Saeedian}
\author[2]{Chengyi Tu}
\author[3]{Fabio Menegazzo}
\author[4]{Paolo D’Odorico}
\author[3]{Sandro Azaele}
\author[3]{Samir Suweis}

\affil[1]{\textit{\normalsize Department of Mathematics, Imperial College London, London, SW7 2AZ, United Kingdom}}

\affil[2]{\textit{\normalsize School of Economics and Management, Zhejiang Sci-Tech University; Hangzhou, 310018, China}}

\affil[3]{\textit{\normalsize Dipartimento di Fisica ``G. Galilei'', Università di Padova, Via Marzolo 8, 35131 Padova, Italy}}

\affil[4]{\textit{\normalsize Department of Environmental Science, Policy, and Management, University of California, Berkeley, CA, USA}}

\begin{document}
\maketitle

\begin{abstract}
Games with environmental feedback have become a crucial area of study across various scientific domains, modelling the dynamic interplay between human decisions and environmental changes, and highlighting the consequences of our choices on natural resources and biodiversity. In this work, we propose a co-evolutionary model for human-environment systems that incorporates the effects of knowledge feedback and social interaction on the sustainability of common pool resources. The model represents consumers as agents who adjust their resource extraction based on the resource's state. These agents are connected through social networks, where links symbolize either affinity or aversion among them. The interplay between social dynamics and resource dynamics is explored, with the system's evolution analyzed across various network topologies and initial conditions.
We find that knowledge feedback can independently sustain common pool resources. However, the impact of social interactions on sustainability is dual-faceted: it can either support or impede sustainability, influenced by the network's connectivity and heterogeneity. A notable finding is the identification of a critical network mean degree, beyond which a depletion/repletion transition parallels an absorbing/active state transition in social dynamics, i.e., individual agents and their connections are/are not prone to being frozen in their social states.
Furthermore, the study examines the evolution of the social network, revealing the emergence of two polarized groups where agents within each community have the same affinity. 
Comparative analyses using Monte-Carlo simulations and rate equations are employed, along with analytical arguments, to reinforce the study's findings. The model successfully captures key aspects of the human-environment system, offering valuable insights to understand how both the spread of information and social dynamics may impact the sustainability of common pool resources.
\end{abstract}

%
%
%
%
%

\section{Introduction}

The sustainable use of natural resources that support the metabolism of human societies is a pressing global challenge \cite{dietz2003struggle, ostrom2008challenge, ostrom2009general}. 
The governance of natural resources that are used in common by multiple individuals (‘common pool resources’, or CPRs) is increasingly recognized as a fundamental challenge to ensure the sustainability of the planet. In the case of open access systems in which resource harvesting is not regulated, users are (relatively) free to over-exploit the CPR \cite{ostrom1990governing, ostrom1994rules, van2007traditions}, a phenomenon known as the “Tragedy of the Commons” \cite{hardin1968tragedy, hardin1998extensions}. In this case, self-interested, short-term, profit-maximizing actors, harvest the resource in disregard of its long-term sustainability. This behavior leads to resource exhaustion, leaving future generations empty-handed \cite{hauser2014cooperating,tu2023emergence}. However, in many cases, CPRs are not open access but the common property of a community; access is restricted to members of the community governing the resources according to internal rules and, in many cases, the sharing of common goals prevents over-exploitation \cite{ostrom1990governing,tu2023emergence}.

From a mathematical perspective, a complex and dynamic interaction between human behavior and CPRs can be modeled as a human-environment system (HES) \cite{ turner2003illustrating, horan2011managing,bauch2016early}. 

In modeling the HES dynamics, one can simplify the problem by considering only the interactions between consumers and resources, neglecting the social structure or resource heterogeneity. In this framework, consumers are treated as players in a game whose payoffs depend on their extraction rates.
Consumer strategies evolve according to game-theoretic principles, based on the rewards or costs of their choices relative to other players (or with respect to the average payoff) \cite{myerson2013fundamentals}. 
It is well known that in non-repeated games, under the assumption of rational and purely self-interested agents, players have an incentive to defect, and defection is the only (Nash) equilibrium. This is a mathematical analog of the tragedy of the commons \cite{myerson2013fundamentals}.

However, to understand the evolution of the HES, it is necessary to consider not only the dynamics of the consumers \cite{hammerstein1994game, hoffman2015robot, cason2014cycles}, but also the dynamics of the resource \cite{sugiarto2015complex,tu2023emergence}. This can be achieved by using one or more differential equations that describe how the resource changes over time, taking into account natural growth, trade, and consumption \cite{tu2023emergence}.
Moreover, the dynamics of consumers and of resources are intrinsically coupled. In fact, there exists a cyclical and responsive relationship between human actions and ecological changes. 
Thus, we need a modeling framework in which players' decisions directly influence the environmental system, and in turn, the evolving state of the environment affects future choices and outcomes in the game \cite{weitz2016oscillating,lin2019spatial,tilman2020evolutionary,tu2023emergence}. 
Furthermore, in this context, the structure of the social network of consumers \cite{chung2013influence} and stochastic fluctuations in resource dynamics \cite{sugiarto2015complex} may also have an important effect on the sustainability of CPRs. However, such elements have not yet been considered in a unified theoretical framework.

In this work, we fill this gap by introducing in the framework of games with environmental feedback three novel elements: 1) users' knowledge feedback, i.e. the behaviour of the player is directly affected by the state of the resources; 2) The consumer social network structure and dynamics, and in particular the effect of polarization in the CPR sustainability; 3)The effect of stochasticity in the resource evolution. 

\begin{figure}[h!]\centering
	\includegraphics[width=1\linewidth]{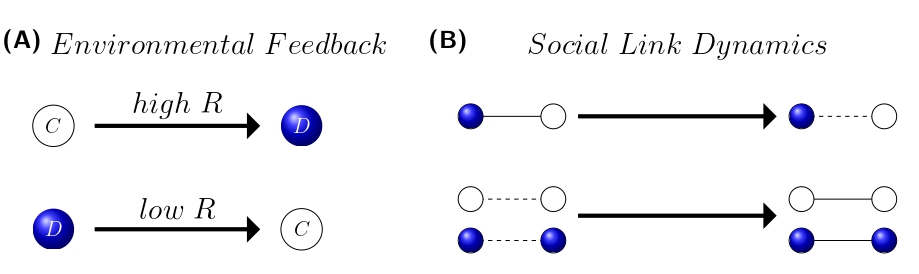}
	\caption{\textbf{Knowledge Feedback: consumers’ adaptation to environmental fluctuations}. (A) Cooperators (white) are more likely to change their behavior in resource-rich environments, while defectors (blue) do so in resource-poor environments. (B) The consumers with different strategies are shown interacting in binary states: positive (cooperators) and negative (defectors). The "social interactions" can have two types of "social link" states: positive (solid) or negative (dashed). The combination of two nodes and connecting link is termed pair-connection, and the sign of each pair-connection is determined by multiplying the sign of constitute components. Three negative pair-connection on the left tend to become positive pair-connection on the right to increase the consumers’ satisfaction.}
	\label{schematic}
\end{figure}

We term knowledge feedback the agents' behavior of adjusting their extraction levels according to the resource state. This means, as documented and studied \cite{schluter2016robustness,dubois2015social,weitz2016oscillating}, that consumers defect when the resource is abundant and cooperate when it is scarce (Fig.~\ref{schematic} (A)).This is rationalized as follows: the classic replicator equation determines the strategy of a player based on the average fitness of the other agents, regardless of the level of resources available. However, we embed the very optimistic assumption that humans will respond to scarce resources by cooperating more.

In our model, we also introduce "social relations" among consumers. These relations are characterized as either \textit{positive/negative links}, symbolizing various types of relationships such as friendship or animosity, trust or distrust, collaboration or intense rivalry, political alignment or opposition, etc. The intensity of the "social interactions" is quantified by assessing the satisfaction level between two interacting individuals, as discussed in \cite{saeedian2019absorbing,saeedian2020absorbing}. It is important to note that the dynamics of "social interaction" are not solely dependent on these social links; the individual states of the agents, like their opinions or strategies, also play a crucial role. For example, a friendship between individuals with conflicting opinions may not endure, leading to dissatisfaction: either they find common ground and align their views, or their relationship deteriorates. Hence, the "social interaction" state is a result of both the individual states of the agents and the nature of their connection (referred to as pair-connection). In our framework, enduring or transient social interactions are represented as \textit{ positive / negative interactions}, respectively. These interactions are also understood as satisfying or unsatisfying \cite{saeedian2019absorbing,saeedian2020absorbing}, and in this work, we use these two terms interchangeably.


One of the main results of our work is that knowledge feedback alone can sustain CPRs in the mean-field case, where consumers ignore the social structure and each agent interacts with all others. In other words, we find that long-term exploitability is achievable if consumers align their behavior with the resource level. However, when social dynamics are also taken into account, they tend to split consumers into two groups sharing the same behavior, leading to a depletion/repletion transition of the CPR, where the stationary resource level and cooperative fraction among agents depend on the average network degree. In particular, we find that hubs in the consumer network favor defectors and cause the depletion phase.

\section{Method}
\subsection{Resource Dynamics}
We model the CPR dynamics using a logistic function that also considers the extraction effort of defectors and cooperators as
\begin{eqnarray}
	\frac{dR}{dt}=TR(1-\frac{R}{K})-R\Big(x \hat{e}_C+(1-x)\hat{e}_D\Big), \quad 0\leq \hat{e}_C<1\leq \hat{e}_D
	\label{resource_dynamics}
\end{eqnarray}
where $T>0$ is the natural growth rate, $K>0$ is the carrying capacity and $x$ is the fraction of cooperators in the systems (that evolve in time, see below). 
The extraction rates of a cooperator and a defector are denoted by $e_C$ and $e_D$, respectively. If $N$ is the number of agents in the systems, we define $\hat{e}_C=N\cdot e_C$, $\hat{e}_D=N\cdot e_D$ and assume $\hat{e}_C\leq \hat{e}_D$, reflecting the fact that cooperators extract fewer resources than defectors \cite{tu2023emergence}.  

It is possible to gain a qualitative understanding of the resource dynamics by rewriting Eq.~(\ref{resource_dynamics}) in terms of a more familiar form of logistic function   
\begin{eqnarray}
	\frac{dR}{dt}&=&T'R\Big(1-\frac{R}{K'}\Big),
\end{eqnarray}
where the growth rate, $T'$, and carrying capacity, $K'$, both depend on $x(t)$, i.e.,
\begin{eqnarray}
	T'=\Big(1-\frac{x \hat{e}_C+(1-x)\hat{e}_D}{T}\Big)T, \quad
 K'=\Big(1-\frac{x \hat{e}_C+(1-x)\hat{e}_D}{T}\Big)K. \nonumber
\end{eqnarray}

Depending on $x(t)$ ($0\leq x(t)\leq 1$), the time-dependent growth rate $T'$ can be positive, zero, or negative,  but it is always bounded between an upper and lower limit, $(T-\hat{e}_D)<T'(t)<(T-\hat{e}_C)$. For everlasting positive growth rates, $T'(0<t<\infty)>0$, the evolution of the resource follows a logistic dynamics, while for stable negative growth rates, $T'(0<t<\infty)<0$, the trajectory of $R$ approaches zero. Moreover, possible scenarios involve oscillations between positive and negative growth rates, leading to oscillations of $R(t)$ between the aforementioned dynamical states. Therefore, if $R$ (at stationarity) approaches zero, then the CPR is in a depleted phase; otherwise it is in a repleted phase. 

The rescaled carrying capacity is bound by upper and lower limits, $(1-\frac{\hat{e}_D}{T})K<K'(t)<(1-\frac{\hat{e}_C}{T})K$. Also, we introduce a rescaled level of available resources $R'$, so we can interpret it as a probability, i.e.,
\begin{eqnarray}
	R’=\frac{R}{(1-\frac{\hat{e}_C}{T})K}, \quad \longrightarrow 0<R’\leq 1
	\label{resource_dynamics_norm}
\end{eqnarray}
Additionally, for simplicity, we set $T=K=1$.

\subsection{Consumers Strategy Dynamics}
\begin{figure}[h!]\centering
	\includegraphics[width=1\linewidth]{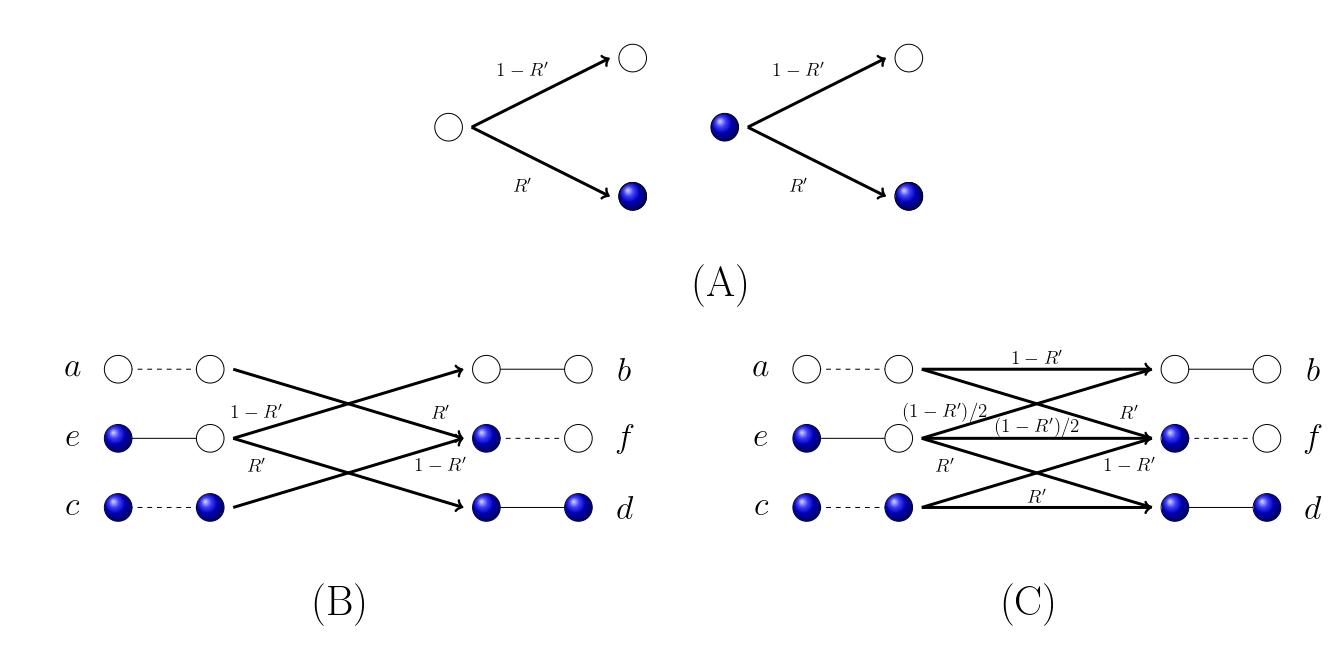}
	\caption{Three models that capture different complex human-environmental dynamics. The white (blue) circles represent cooperator (defector) and solid (dash) links represent positive (negative) connections.  (A) A unified update rule is derived from the node dynamics of knowledge feedback without interaction (\textbf{NS}). (B) A unified update rule is derived from the interplay between the knowledge feedback and the quenched binary-state interactions, which we call (\textbf{QBS}). (C) A unified update rule results from the interplay between the node dynamics of knowledge feedback and the evolving binary-state social links, which we call (\textbf{EBS}). The pair-connections $a$, $c$ and $e$ denote negative social interactions, while the pairs $b$, $d$ and $f$ indicate positive social interactions. We illustrate here the social dynamical rules that transform nanegative interactions into positive interactions through node or link updates. $R^{\prime}$ is the normalized resource level. In this work, we have implemented a sequential update scheme. }
	\label{update_rules}
\end{figure}

The consumers' strategy dynamics, i.e. how the fraction of cooperators $x$ evolves in time, is governed by the update rules graphically presented in Fig.~\ref{update_rules}. 

We investigate three different scenarios of increasing complexity, corresponding to different game update rules driving the evolution of the users' social networks and of the HES.

First, we study the stochastic dynamics of the HES when only knowledge feedback is considered and no social structure (\textbf{NS}) is included, as depicted in Fig.~\ref{update_rules} (A).

According to this model, the consumers’ strategies depend only on the availability of resources: they tend to cooperate when resources are scarce and to defect when resources are plentiful.
To implement this model, we make two fundamental assumptions: 1) the probability of switching from cooperation to defection is linearly proportional to the normalized resource level  ( $R^{\prime}\in [0,1]$ ); 2) The probability of switching from defection to cooperation is linearly proportional to the complementary normalized resource level ( $1-R^{\prime}$ ).

Then we focus on the essential role of complex social interactions \cite{castellano2009statistical} between consumers. 
We assume that social connections (link) can be either positive or negative and that they can be quenched (\textbf{QBS}) or evolve in time (\textbf{EBS}) \cite{saeedian2019absorbing,saeedian2020absorbing} as shown in Figs.~\ref{update_rules} (B)-(C). 
Regardless of state of nodes, the positive/negative connection (link) can be interpreted in different ways,  friendship or animosity, trust or distrust, collaboration or intense
rivalry, political alignment or opposition, etc., depending on the context and the nature of the problem. 
We note that many complex systems exhibit binary connections, similar to attraction/repulsion in physics, friendship/enmity in opinion formation \cite{saeedian2019absorbing,saeedian2020absorbing,antal2005dynamics}, mutualism/competition in ecology \cite{allesina2012stability,suweis2013emergence}, or excitatory/inhibitory in neuroscience \cite{apicella2021emergence}.

For the \textbf{QBS} and \textbf{EBS} case, we propose a dynamical update rule based on the interaction between binary state nodes (defectors / cooperators) and binary state links (positive / negative), as illustrated in Fig.~\ref{update_rules} (B-C). The negative pair-connections $\{a,c,e\}$ correspond to short-lasting social interactions and tend to evolve into long-lasting ones $\{b,d,f\}$ to maximize the level of satisfaction irreversibly. 
This condition breaks the detailed balance condition and drives the dynamics out of equilibrium.

To identify positive / negative pair connections, we can adopt the energy function approach for signed social networks \cite{facchetti2011computing,singh2014extreme,kargaran2021heider,minh2020effect}. In this approach, the unbiased binary variable ($\pm$) represents the node and link's state. The main principle behind this approach is that the negative pair-connections in social systems are characterized by an odd number of negative nodes and links, whereas the positive pair-connections are characterized by an even number of negative nodes and links. To be more precise, the sign (positive or negative) of a pair-connection is determined by multiplying the signs of the three components involved in that pair-connection. For example, a configuration of two defectors with a negative link constitutes a configuration with three negative components, which results in a negative pair-connection $c$. These three-component interactions are known as higher-order interactions.



We present in Appendix.~\ref{Appendix_neg_pos} the interpretations and the respective roles of each of the six possible pair-connections $\{a,b,c,d,e,f\}$ in the update rule, where pair-connections $\{a,c,e\}$ are positive (socially satisfying) and $\{b,d,f\}$ are negative (socially unsatisfying) interactions.

\subsection{Coupled HES Dynamics}
We thus propose a co-evolutionary model for HES that incorporates two interdependent dynamics of human and environment sub-systems, where the density of cooperators in the human sub-system $x$ and the level of resource in the environment sub-system $R$ co-evolve.

Putting all the pieces together, the HES dynamics can be modelled by an undirected network of $N$ individuals, where the nodes represent the consumers with binary state extraction strategies (cooperator/defector) and the links represent the binary state social connections. 
The density of pair-connections in the network is given by $\rho_i=\frac{L_i}{L}, i \in \{a,b,c,d,e,f\}$, where $L_i$ is the number of pair-connections $i$ and $L$ is the total number of links in the network.
 

To investigate these co-evolving dynamics we use two methods: microscopic Monte-Carlo simulations and macroscopic rate equations. 
We explain these methods in detail in the following subsections.

\subsection{Monte-Carlo method}

We use Monte-Carlo simulations to study the co-evolutionary dynamics of update rules (Fig.~\ref{update_rules}) and resource dynamics (Eq.~\ref{resource_dynamics}) on complex networks, mainly Erd\H{o}s-R\'enyi networks with size $N$ and mean degree $\mu$, unless otherwise stated.
A Monte Carlo step consists of a sequence of events, each involving one constituent of the system.
The constituents are either nodes (in the case of knowledge feedback without interaction, Fig.~\ref{update_rules} (A)) or pair-connections (in the case of knowledge feedback with interaction, Fig.~\ref{update_rules} (B) and (C)).
Thus, one computer iteration corresponds to $\frac{1}{N}$ or $\frac{1}{L}$ Monte-Carlo time steps, depending on the case.

\begin{itemize}
	\item \textbf{NS} model: We use a Monte-Carlo method to simulate the dynamics shown in Fig.~\ref{update_rules} (A), where at each iteration, a randomly selected node updates its state according to a randomly generated number $\xi$ between $0$ and $1$. The node becomes a cooperator if $R^{\prime}<\xi$, or a defector otherwise. The initial strategy of the node is irrelevant.
	\item \textbf{QBS} and \textbf{EBS} models: We use a Monte-Carlo method to implement the dynamics shown in Fig.~\ref{update_rules} (B) and (C), where at each iteration, a randomly selected pair-connection updates its state according to the corresponding update rules if it belongs to one of the negative interactions $\{a,c,e\}$. Otherwise, no update occurs. For example, in the \textbf{QBS} dynamics (Fig. ~\ref{update_rules}(B)), if a pair-connection $e$ is chosen and a random number $\xi$ is generated, the cooperator switches to defection if $R^{\prime}>\xi$, or the defector switches to cooperation otherwise. However, in the EBS dynamics (Fig. ~\ref{update_rules}(C)), if a pair-connection $e$ is chosen, the process works slightly differently. Similarly, the cooperator switches to defection if $R^{\prime}>\xi$. But for the complementary case, instead of the defector simply switching to cooperation, there are two possible outcomes with equal probability: either the defector switches to cooperation, or the positive link turns into a negative link.
	\item Resource dynamics: Along with the node and pair-connection updates, the resource level also changes at each iteration. Using the Euler method and the updated value of $x$, we can advance the resource dynamics as follows $R(t+\Delta t) = R(t)+\Delta t\Bigg(R(t)(1-R(t))-R(t)\Big(\hat{e}_Cx(t)+\hat{e}_D(1-x(t))\Big)\Bigg)$, where for the case without interaction (Fig.~\ref{update_rules} (A)), we have $\Delta t=\frac{1}{N}$, and for the case with interaction (Fig.~\ref{update_rules} (B) and (C)), we have $\Delta t=\frac{1}{L}$.
\end{itemize}

\subsection{Macroscopic rate equations method}

We can model the temporal evolution of the pair-connection densities in the social dynamics (Fig.~\ref{update_rules} (B) and (C)) as a multi-variable death-birth Markov process with a master equation (Appendix.~\ref{Appendix_ME_RE}). 
However, the resource dynamics only depend on the cooperator density $x$ (Eq.~\ref{resource_dynamics}), not on the pair-connections. 
To address this, we can derive an explicit relation between the pair-connection densities and the cooperator density for a random homogeneous network as follows (Appendix.~\ref{Appendix_x_rho}):
\begin{eqnarray}
	x&=&\rho_{a}+\rho_{b}+\frac{\rho_{e}+\rho_{f}}{2}  \cr
	\phi&=&\rho_{c}+\rho_{d}+\frac{\rho_{e}+\rho_{f}}{2},
	\label{x_pair}
\end{eqnarray}
where $x$ and $\phi$ are the cooperator and defector densities, respectively, and $x+\phi
=1$. This relation also applies to an Erd\H{o}s-R\'enyi network when $N\gg1$ (Appendix.~\ref{Appendix_x_rho}). 
Using this approach, we can obtain the deterministic rate equations for the temporal evolution of pair-connection densities for the evolving dynamicl binary state interaction dynamics (Fig.~\ref{update_rules} (C)) as follows
\begin{eqnarray}
\	\frac{d\rho_{a}}{dt}&=&-\rho_{a}+(\mu-1)\Bigg[-\frac{R'}{x}\Big(\rho_{a}+\rho_{e}\Big)\rho_{a}+\frac{(1-R')}{2\phi}\Big(\rho_{c}+\frac{\rho_{e}}{2}\Big)\rho_{f}\Bigg]  \cr
	\
	\frac{d\rho_{b}}{dt}&=& (1-R')\rho_{a}+\frac{(1-R')}{2}\rho_{e}+(\mu-1)\Bigg[-\frac{R'}{x}\Big(\rho_{a}+\rho_{e}\Big)\rho_{b}+\frac{(1-R')}{2\phi}\Big(\rho_{c}+\frac{\rho_{e}}{2}\Big)\rho_{e}\Bigg] \cr
	\
	\frac{d\rho_{c}}{dt}&=&-\rho_{c}+(\mu-1)\Bigg[\frac{R'}{2x}\Big(\rho_{a}+\rho_{e}\Big)\rho_{f}-\frac{(1-R')}{\phi}\Big(\rho_{c}+\frac{\rho_{e}}{2}\Big)\rho_{c}\Bigg]  \cr
	\
	\frac{d\rho_{d}}{dt}&=& R'\rho_{c}+R'\rho_{e}+(\mu-1)\Bigg[\frac{R'}{2x}\Big(\rho_{a}+\rho_{e}\Big)\rho_{e}-\frac{(1-R')}{\phi}\Big(\rho_{c}+\frac{\rho_{e}}{2}\Big)\rho_{d}\Bigg]\cr
	\
	\frac{d\rho_{e}}{dt}&=& -\rho_{e}+(\mu-1)\Bigg[\frac{R'}{x}\Big(\rho_{a}+\rho_{e}\Big)\Big(\rho_{b}-\frac{\rho_e}{2}\Big)+\frac{(1-R')}{\phi}\Big(\rho_{c}+\frac{\rho_{e}}{2}\Big)\Big(\rho_{d}-\frac{\rho_e}{2}\Big)\Bigg] \cr
	\
	\frac{d\rho_{f}}{dt}&=&R'\rho_{a}+(1-R')(\rho_{c}+\frac{\rho_{e}}{2})+(\mu-1)\Bigg[\frac{R'}{x}\Big(\rho_{a}+\rho_{e}\Big)\Big(\rho_{a}-\frac{\rho_f}{2}\Big)+   \cr &&\frac{(1-R')}{\phi}\Big(\rho_{c}+\frac{\rho_{e}}{2}\Big)\Big(\rho_{c}-\frac{\rho_f}{2}\Big)\Bigg],
	\label{RE}
\end{eqnarray}
where $\mu$ is the mean degree of the random homogeneous network, and this coupled differential equation depends on normalized level of resource $R^{\prime}$. An analytical relationship between $x$ and $R^{\prime}$ is provided in Eq.~\ref{x_R_mu} when Eq.~\ref{RE} reaches the stationary state. Then, an exact analytical solution for the co-evolution of Eq.\ref{resource_dynamics} coupled with Eq.~\ref{RE} is obtained by computing the roots of Eq.~\ref{R_mu_e_CD}. 


We can also obtain the rate equations for the social dynamics shown in Fig.~\ref{update_rules} (B) by using the same method, as shown in Eq.~\ref{RE_quenched} (Appendix.~\ref{Appendix_ME_RE}).
When the stationary state is reached in Eq.~\ref{RE_quenched}, Eq.~\ref{x_R_mu_q} provides an exact analytical relationship between $x$ and $R^{\prime}$ for an important particular case where there is no negative interaction. Then, an exact analytical solution is provided for the co-evolution of Eq.\ref{resource_dynamics} coupled with Eq.~\ref{RE_quenched} presented in Eq.~\ref{R_mu_e_CD_q}.

All the above results obtained using this approach agree with those obtained from the full integration of the differential rate equations (Eq.~\ref{RE} and Eq.~\ref{RE_quenched} coupled with Eq.\ref{resource_dynamics}).

\section{Results}

We first analyze in detail the dynamics of each of the three model variants (\textbf{NS}, \textbf{QBS}, \textbf{EBS}) on Erd\H{o}s-R\'enyi networks. Then we briefly discuss the main features of the dynamics on Scale-Free and Small-World networks. 

For each realization of the dynamics, we generate a random complex network and randomly assign the initial states of nodes and links. 
We denote by $x_0$ the initial density of cooperative players, by $l_0$ the initial density of positive links, and by $R^{\prime}_0$ the initial value of normalized resource level.

For each HES dynamics, we compare the Monte-Carlo results with the numerical solutions of the rate equations, Eq.~\ref{RE}, and - as we will show - we find a good agreement between them. 

We are interested in the conditions that lead to either a depleted or a repleted phase. 
A repleted phase is a phase where resources are used sustainably (${R'}^{\text{st}}>0$), while a depleted phase is a phase where resources are exhausted (${R'}^{\text{st}}=0$). 
This depletion/repletion transition in resource dynamics also corresponds to a transition from an absorbing to an active state in the temporal evolution of pair-connection densities \cite{saeedian2019absorbing,saeedian2020absorbing}. 
In an absorbing state, the densities of negative pair-connections ($\rho_a$,$\rho_c$ and $\rho_e$) decay exponentially to zero, while in an active state, they reach a non-zero steady state \cite{saeedian2019absorbing,saeedian2020absorbing}. The absorbing/active phase can be interpreted in social terms as an emergent situation where the individuals and their connections are/are not prone to being frozen in their current social
states, respectively.
In the following subsections, we present the results of the co-evolution of Eq.~\ref{resource_dynamics} and the three social dynamics shown in Fig.~\ref{update_rules}.

\subsection{No-social interaction (\textbf{NS}).}

We consider the HES dynamics with knowledge feedback but without any social interaction. 
The temporal evolution of the cooperator density can be written as
\begin{eqnarray}
	\dot{x}&=&-R’x+(1-R’)(1-x).
	\label{x_NS}
\end{eqnarray}
The normalized resource level at steady state in the co-evolution of Eq.~\ref{resource_dynamics} and Eq.~\ref{x_NS} is given by $R^{\prime}(t\rightarrow\infty)={R^{\prime}}^{\text{st}}=\frac{1-\hat{e}_C}{1-2\hat{e}_C+\hat{e}_D}$, which is always positive for any values of $\hat{e}_C$ and $\hat{e}_D$. 
Finally, we note that $x^{\text{st}}=1-{R^{\prime}}^{\text{st}}$ for any value of $\hat{e}_C$ and $\hat{e}_D$, and thus $x^{\text{st}}>0.5$. Therefore, \textbf{NS} systems always lead to cooperators being the majority in a society. In other words, for a sustainable cooperative HES to be achieved, the community of cooperators must always form a majority, $x^{\text{st}}>0.5$. Otherwise, the system is likely to collapse.

\subsection{Quenched binary state interaction (\textbf{QBS}).}

\begin{figure}[h!]\centering
	\includegraphics[width=1\linewidth]{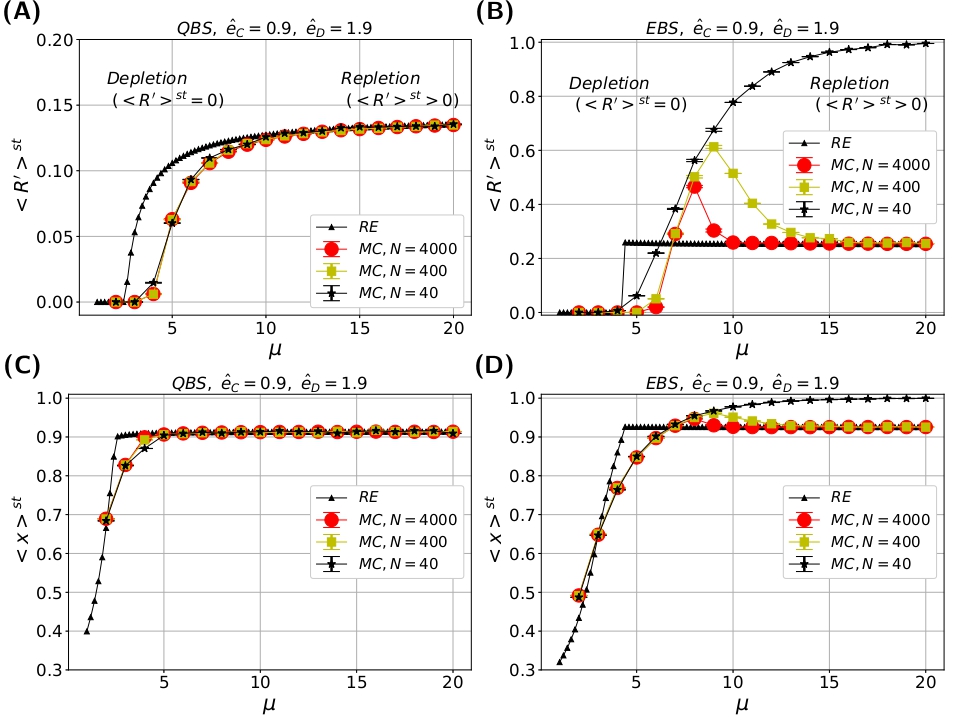}
	\caption{Stationary states of the normalized level of resources ${<R^{\prime}>}^{\text{st}}$ and the density of cooperators $<x>^{\text{st}}$ as functions of network mean degree $\mu$. (A) and (C) The \textbf{QBS} dynamics shows a transition from a depleted state to a repleted state as $\mu$ increases. (B) and (D) The \textbf{EBS} dynamics shows a similar transition. We compare two dynamics with networks of different sizes $N=40, 400, 4000$, at $(\hat{e}_C=0.9,\hat{e}_D=1.9)$. We show in all figures the results from the Monte-Carlo simulation and numerical integration of differential rate equations, labeled by MC and RE. The initial value of the density of cooperators, the fraction of positive links, and the normalization level of the resource are $x_0=0.25$ and $l_0=0.5$, and $R^{\prime}_0=\frac{2}{3}$, respectively (see Appendix.~\ref{Appendix_ic}). The ${<R^{\prime}>}^{\text{st}}$ (in panel (A) and (B)) is insensitive to different initial values, as evident in Fig.~\ref{results_R_initi_R}.}
	\label{results_R}
\end{figure}

We now consider the HES dynamics with knowledge feedback and with quenched binary state interaction, as shown in Fig.~\ref{update_rules}(B). 
We simulate the co-evolution of this social interaction and the resource dynamics on an uncorrelated Erd\H{o}s-R\'enyi network for three different sizes $N=40, 400,4000$. 
We plot the normalized resource level, $R^{\prime}$, and the cooperator density $x$, at steady state, as functions of mean degree, $\mu$, in Fig.~\ref{results_R}(A) and (C), respectively. 
We also show the corresponding curves obtained from numerically solving the rate equations Eq.~\ref{RE_quenched}. 
Both Monte-Carlo and the rate equations simulations indicate that increasing $\mu$, which acts a control parameter, leads to a transition from a depleted to a repleted state. In fact, in the limit $\mu \rightarrow N-1$, we qualitatively recover the pure knowledge feedback case without social structure (see Appendix.~\ref{Appendix_in_so}).
There is a good agreement between the results of the rate equations and of the Monte-Carlo simulation, especially for larger $\mu$ when the Erd\H{o}s-R\'enyi network becomes more homogeneous. 
Near the critical point, both Monte-Carlo and rate equations results show qualitative agreement.

\begin{figure}[h!]\centering
	\includegraphics[width=1\linewidth]{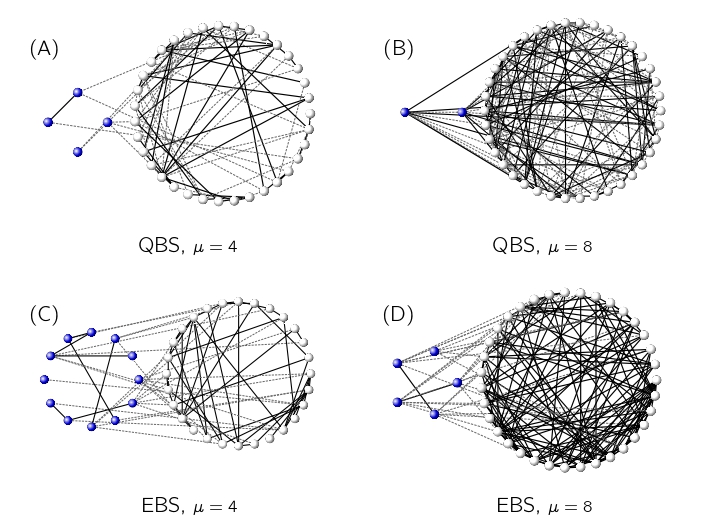}
	\caption{(A)\&(B) Jammed states of the quenched binary state dynamics (\textbf{QBS}) and (C)\&(D) examples of absorbing configurations of the evolving binary state dynamics (\textbf{EBS}) on a finite system with extraction coefficients ($\hat{e}_C=0.9,\hat{e}_D=1.9$). The network has $N=40$ nodes and mean degree $\mu=4$ and $\mu=8$. The initial values are $x_0=0.25$, $l_0=0.5$ (see Appendix.~\ref{Appendix_ic}), and $R^{\prime}_0=\frac{2}{3}$. For both case of \textbf{QBS} and \textbf{EBS}, the fraction of defectors (blue circles) decreases as the network connectivity increases.}
	\label{QBS_EBS_plot}
\end{figure}

In a finite-size system, for any value of $\mu$, the system may enter a frustrated jammed state \cite{antal2005dynamics,saeedian2017epidemic,marvel2009energy}, where the quenched binary state interactions prevent reaching an absorbing state. 
In this configuration, we observe a coexistence of negative ($a$,$c$,$e$) and positive ($b$,$d$,$f$) pair-connections, although the system is not dynamically active, as shown in Fig.~\ref{QBS_EBS_plot} (A)\&(B). 
In other words, applying the update rule on a jammed state does not change the arrangement of node states in the system, except for some rare possible node-blinking events. 
In the structural balance theory, the probability of reaching a jammed state in fully connected network decreases rapidly as $N$ increases \cite{antal2005dynamics}. We observed a similar behaviour in our system for large $\mu$. However, examining the probability of the jammed state in sparse networks is not within the scope of this study.
Finally, as evident from Fig.~\ref{results_R}(A) and (C), increasing network connectivity improves the sustainability of the HES. This is true even in the emergence of jammed state in the finite size systems.

\subsection{Evolving dynamical binary state interaction (\textbf{EBS})}

We now study the phase structure of the co-evolution of the complex social interaction shown in Fig.~\ref{update_rules}(C) with resource dynamics given by Eq.~\ref{resource_dynamics}. 
We plot the normalized resource level, $R^{\prime}$, and the cooperator density $x$, at steady state, as functions of mean degree, $\mu$, for three different sizes $N=40, 400, 4000$ in Fig.~\ref{results_R} (B) and (D), respectively. 
Again, in both the Monte-Carlo simulations and rate equations of the system, for any (large enough) sizes, we show the existence of a critical mean-degree $\mu_c$ below which a depleted state occurs. 
In Monte Carlo simulations, for large sizes ($N=400, 4000$), above $\mu_c$, the curves overshoot (reach a maximum value, and for larger $\mu$ they approach a plateau), as seen in Fig.~\ref{results_R}(B) and (D). 
On the other hand, in the curves obtained from the rate equations the maximum occurs at $\mu_c$ discontinuously, as shown by black-triangle markers.
Interestingly, for larger $\mu$ when the Erd\H{o}s-R\'enyi network becomes more homogeneous, rate equations and Monte-Carlo results are very consistent. 
For small size ($N=40$), we observe a sigmoid-like behavior for $R^{\prime}$, from a depleted phase to a fully repleted phase, as indicated by black-star marker in Fig.~\ref{results_R}(B).

\begin{figure}[h!]\centering
	\includegraphics[width=1\linewidth]{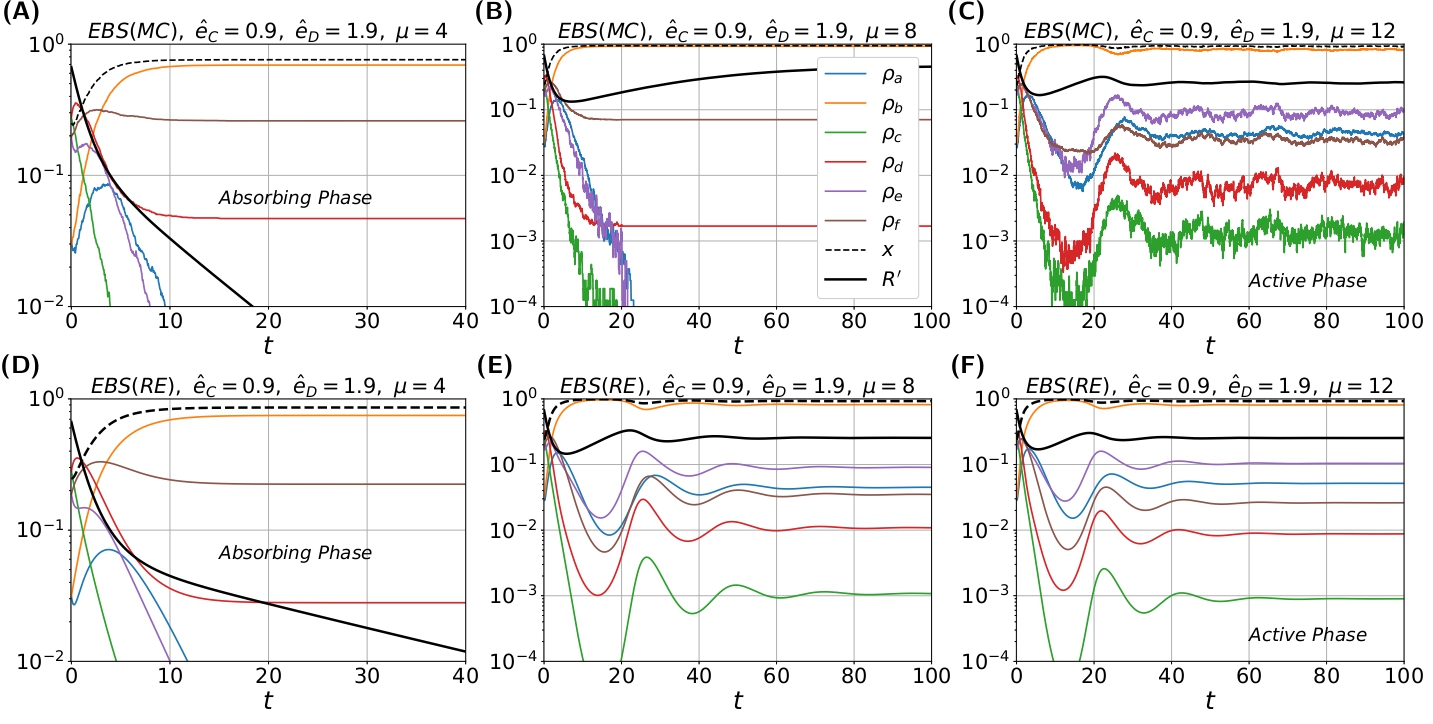}
	\caption{The temporal evolution of normalized resource level, $R^{\prime}$, cooperator density, $x$, and pair-connection densities $\rho_i$, for evolving dynamical binary interaction (\textbf{EBS}). The top panels show the results of the Monte-Carlo simulation of a single run on an Erd\H{o}s-R\'enyi network with $N=4000$. The bottom panels show the results of the corresponding rate equations. Panels (A) and (D) represent the parameters leading to an absorbing phase, while panels (C) and (F) represent the parameters leading to an active phase. Panels (B) and (E) represent the parameters leading to the critical region where the value of ${R^{\prime}}^{st}$ in panel (B) of Fig.~\ref{results_R} reaches the maximum value. The initial values for all dynamics are $x_0=0.25$, $l_0=0.5$ (see Appendix.~\ref{Appendix_ic}), and $R^{\prime}_0=\frac{2}{3}$.}
	\label{Trajectories}
\end{figure}

The depletion/repletion phase transition observed in Fig.~\ref{results_R} (B) and (D) is accompanied by a non-equilibrium absorbing-state transition \cite{saeedian2019absorbing,saeedian2020absorbing}. By definition, an absorbing state is a configuration that the model can reach but from which it cannot escape. In our case, the absorbing state is identified by a solution when $\rho_a=\rho_c=\rho_e=0$. In general, in the absorbing phase, the absorbing state is a stable solution against small perturbations, while in the active phase, the absorbing state becomes unstable (see equation 121 of \cite{hinrichsen2000non}).
To characterize this transition, we monitored the trajectories of $R^{\prime}$, $x$, and $\rho_i$ for three sets of parameters leading to either an absorbing or an active phase. 
Fig.~\ref{Trajectories}(A) and (D) represent the dynamics in an absorbing phase, while (C) and (F) represent an active phase. 
The middle panels represent the trajectories of the system for a set of parameters leading to the maximum value of $R^{\prime}$ (close to the critical point), as seen in panel (B) of Fig.~\ref{results_R}. 
As shown, in an absorbing phase, the resource level $R^{\prime}$ and the densities of negative pair-connections ($a,c,e$) decay exponentially to zero, while in an active phase they reach nonzero steady states. 
However, in the upper middle panel, although the dynamics indicate an absorbing phase, $R^{\prime}$ reaches a non-zero plateau. We speculate that this is due to a strong fluctuating behavior close to criticality, and related finite size effects. 
The top panels are obtained from Monte-Carlo simulations, while the bottom panels are obtained from rate equations with the same set of parameters. 
As shown, except for the middle panels, the Monte-Carlo results agree well with the corresponding rate equations results.

A snapshot of an absorbing configuration for the evolving binary state dynamics (\textbf{EBS}) on a finite system is shown in Fig.~\ref{QBS_EBS_plot} (C)\&(D). 
As seen in this figure, the fraction of defectors decreases with increasing network connectivity (Fig.~\ref{results_R}(D)), revealing the underlying mechanism behind the depletion/repletion phase transition in finite systems.
In the absorbing phase, the average characteristic time to reach an absorbing state grows logarithmically with system size ($<\tau>\sim \log{N}$), while in the active phase, it grows exponentially ($<\tau>\sim \exp{N}$), as shown in Appendix.~\ref{Appendix_tau} and Fig.~\ref{tau}. 
Finally, as evident from panels (B) and (D) of Fig.~\ref{results_R}, increasing the network connectivity enhances the sustainability of the HES.


\subsection{Robustness of Results.}


We have shown that the co-evolution dynamics on Erd\H{o}s-R\'enyi networks, with relatively symmetric initial conditions, can lead to a transition from depletion to repletion of CPRs, depending on the network connectivity. In particular, we have shown that the higher mean degree of the Erd\H{o}s-R\'enyi system promotes CPR sustainability. 

Our results are robust by varying initial conditions, the network topology, and the social model (the similarity of the main results under both the \textbf{EBS} and \textbf{QBS} models).

Fig.~\ref{results_R_initi_R} shows the co-evolution dynamics of ${<R^{\prime}>}^{st}$ and $<x>^{st}$ as a function of mean degree for the model in Fig.\ref{update_rules}(C). The Monte-Carlo and rate equations simulations were initialized in a wide range of combinations of $R'_0$ and $x_0$ and run on Erd\H{o}s-R\'enyi networks. Regardless of the initial conditions, ${R'}^{st}$ evolves toward the same stationary state dictated by the mean degree. This suggests the co-evolution rules drive the CPR to a robust equilibrium that depends only on the mean degree, erasing memory of initial seeds $R'_0$ and $x_0$. Universality across different initial configurations indicates that the stationary state is an attractor for the resource dynamics on Erd\H{o}s-R\'enyi networks. Further analyses of the phase space are presented in (Appendix.~\ref{Appendix_IC}\&Appendix.~\ref{Appendix_e_D_e_C}).

In Fig.~\ref{results_R_initi}, we also plot ${<R^{\prime}>}^{st}$ and $<x>^{st}$ as functions of network mean degree for four initial cooperator densities ($x_0 = 0.05, 0.25, 0.75, 0.95$), and three network topologies, Erd\H{o}s-R\'enyi, Barabasi-Albert \cite{barabasi1999emergence} and Watts-Strogatz \cite{watts1998collective}. 
We find that the depletion/repletion transition is qualitatively similar for different networks heterogeneities.
However, the Barabasi-Albert network, which has higher heterogeneity, exhibits a shift of the critical point for higher average degree.
This implies that the hubs in the network facilitate the depletion phase.
Another notable observation is that, in the depletion phase, the resource level reaches a depleted state of ${<R^{\prime}>}^{st}=0$, regardless of the initial condition of the dynamics (absorbing phase in  Fig.\ref{results_R_initi} in panels (D), (E), and (F)) or the network topology (absorbing phase in Fig.\ref{results_R_initi} panels (A), (B), and (C)). 
This indicates that the depletion/repletion transition on the complex networks are robust to initial conditions.

To vary the social model, we already study the \textbf{EBS}, presented in Fig.~\ref{update_rules}(C), and its quenched version,  \textbf{QBS}, presented in Fig.~\ref{update_rules}(B). As shown in Fig.~\ref{results_R} (A) and (B), the depletion/repletion emerges in both the \textbf{EBS} and \textbf{QBS} models. This suggests that the characteristic signature of the transition is robust and does not rely directly on the specific dynamics of the link. However, the types of transitions in  \textbf{EBS} and \textbf{QBS} are different (see Appendix.~\ref{Appendix_Analytic}). As shown in Fig.~\ref{results_R} (A), Monte-Carlo  and rate equations simulations for the \textbf{QBS} model both indicate the depletion/repletion transition is continuous (see also the analytical arguments in Appendix.~\ref{Appendix_Analytic_QBS}), while,
describing the type of transition in \textbf{EBS} (Fig.~\ref{results_R} (B)) is a challenging task that we leave for future work. In Appendix.~\ref{Appendix_Analytic_EBS}, we provide some analytical arguments that seem to indicate that the rate equations, differently from the Monte-Carlo simulations, show a discontinuous phase transition. Also in (Appendix.~\ref{Appendix_e_D_e_C}) we provide some remarks about the critical line of the depletion/repletion transition in the phase space of  ($\hat{e}_C$,$\hat{e}_D$).  

Studying the \textbf{QBS} offers additional insights beyond \textbf{EBS} alone. The \textbf{QBS} retain most of the key characteristics of the full socio-environmental dynamics. However, they exhibit a more well-behaved continuous phase transition between the active and absorbing states, rather than the unusual punctuated transition observed in the \textbf{EBS} (compare panel (A) and (B) in Fig.~\ref{results_R}). Analyzing this continuous phase transition in the \textbf{QBS} may lend itself more easily to adapting standard physics frameworks to understand phase transitions. Thus, the \textbf{QBS} analysis complements the \textbf{EBS} picture while facilitating connection to established theory.

Finally, it is worth mentioning that the dynamics of \textbf{QBS} and \textbf{EBS} in the network with a high mean degree behave similarly to \textbf{NS} dynamics. The time evolution of the density of cooperators $x$ in \textbf{NS}, \textbf{QBS}, and \textbf{EBS} can be represented mathematically, respectively, as follows:
\begin{eqnarray}
	\dot{x}&=&-R’x+(1-R’)(1-x),
	\label{x_NS_0}
\end{eqnarray}
\begin{eqnarray}
	\dot{x}&=&\mu\Big(-R'(\frac{\rho_{a}+\rho_{e}}{2})+(1-R')(\frac{\rho_{c}+\rho_{e}}{2})\Big),
\label{x_derivative_q_0}
\end{eqnarray}
\begin{eqnarray}
	\dot{x}&=&\mu\Big(-R'(\frac{\rho_{a}+\rho_{e}}{2})+(1-R')(\frac{\rho_{c}+\frac{\rho_{e}}{2}}{2})\Big).
\label{x_derivative_0}
\end{eqnarray}
The derivation of Eq.\ref{x_derivative_q_0} and Eq.\ref{x_derivative_0} are presented in Appendix.~\ref{Appendix_Analytic} in Eq.\ref{x_derivative_q} and Eq.\ref{x_derivative}, respectively. The Eq.\ref{x_NS_0} on one side and the Eqs.\ref{x_derivative_q_0}-\ref{x_derivative_0} on the other side have some similarity. As is evident from Eq.\ref{x_pair}, $\rho_a$ and $\rho_e$ are constituent components of $x$, while $\rho_c$ and $\rho_e$ are constituent components of $1-x$. As fully clarified in Appendix.~\ref{Appendix_in_so}, roughly speaking, for larger $\mu$ we expect a higher activity of the system, which results in a higher value of negative pair-connections ($\rho_a$, $\rho_c$, and $\rho_e$) and, in turn, makes Eqs.\ref{x_derivative_q_0}-\ref{x_derivative_0} more similar to Eq.\ref{x_NS_0}.


\subsection{Polarization of the Social Network Dynamics}
Quantifying the degree of social polarization is a key issue \cite{shekatkar2018importance,hohmann2023quantifying,saeedian2019absorbing,saeedian2020absorbing} we have tackled in this work.
A major challenge here is to devise a suitable measure for polarization.
We propose one that is aligned with the aim of this study.
We contend that polarization between two groups should depend not only on their different strategies but also on their antagonism towards each other.
Polarization arises only when the inter-group connection are negative due to the different strategies (which is not a trivial condition).
For instance, if $x=0.5$, but all the links between the groups are positive, then there is no meaningful polarization (which means that the groups cooperate despite having different opinions).

Hence, we define the polarization index, $r$, as the product of the polarization due to the node-attribute (Eq.~\ref{x_pair_rel}), $2(\rho_f + \rho_e)$ (which is equal to equation 3 in \cite{shekatkar2018importance}), and the polarization due to the link-attribute ( $\frac{\rho_f}{\rho_f+\rho_e}$ ).
Therefore, the polarization index has the simple form $r=2\rho_f$.
This is interesting because $\rho_f$ depends on initial conditions in the depletion phase, but not in the sustainable phase.

The network polarization is shown in Fig.~\ref{results_R_initi}(G), (H) and (I) for the phase space of ( $\hat{e}_C=0.9,\hat{e}_D=1.9,\mu$ ).
We observe that the depletion phases have higher polarization than the repletion phases, indicating that lower network connectivity leads to higher polarization.
This agrees with our intuition.
However, we also find that lower extraction coefficients ( $\hat{e}_C,\hat{e}_D$ ) result in higher polarization (as evident in Fig.~\ref{results_R_mu_fiex}(C)), which is counter-intuitive (See Appendix.~\ref{Appendix_e_D_e_C}).

In all model configurations (\textbf{NS}, \textbf{QBS}, and \textbf{EBS}), and in both the repleted and depleted phases, the majority of the consumers deploy a cooperation strategy (to be quantitative, $x^{\text{st}}>0.5$). In fact, the dynamics of \textbf{NS} always leads to $x^{\text{st}}>0.5$. On the other hand, the diagram of $x^{\text{st}}$ in terms of $\mu$ for \textbf{QBS}, and \textbf{EBS} are presented in Fig.~\ref{results_R} (C) and (D) respectively. As evident, in the depleted phases, for $\mu>2$, the cooperators build the majority, and in the repleted phase, both systems must achieve a majority in excess of 90 percent cooperation (the overwhelming majority). In particular, in the \textbf{EBS} case, Fig.~\ref{results_R_mu_fiex} (B)\&(E) shows $x^{\text{st}}$ as a function of ($\hat{e}_C,\hat{e}_D$). It is interesting to note that depleted phases tend to have the highest cooperator density, $x^{\text{st}}>0.7$ (the depleted phase is identified by the black region in panel (A)\&(D)). In other words, the HES system may still not be sustainable even if the majority of extractors are cooperators.

A likely explanation for the asymmetric behavior of $x^{st}$ is the inherent asymmetry in the extraction terms in Eq. \ref{resource_dynamics}, where the cooperators' extraction capacity $\hat{e}_C$ is less than the defectors' extraction capacity $\hat{e}_D$. Specifically, the form of the extraction terms $\hat{e}_Cx$ and $\hat{e}_D(1-x)$ suggests that they contribute comparably to the total resource dynamics despite the inequality $\hat{e}_C<\hat{e}_D$. To compensate, some underlying driver in the system skews the dynamics toward higher steady-state cooperator densities $x^{st}>(1-x^{st})$. Through this skew, the lower extraction capacity of cooperators is counterbalanced by their increased representation, leading $\hat{e}_Cx$ to be approximately equal to $\hat{e}_D(1-x)$ in equilibrium. The system appears to rebalance the asymmetry in the extraction terms by shifting the stable ratio of cooperators to defectors.

\section{Discussion and Conclusions}

In this work, we propose a co-evolutionary model for human-environment systems (HESs) that incorporates the effects of knowledge feedback and users' social interaction dynamics on the sustainability of common pool resources (CPRs). 

The proposed framework allows us to capture the complex and dynamic interplay between human behaviors and resource availability and to explore how different scenarios affect the sustainability of CPRs. In particular, we have found that knowledge feedback alone, when fluctuations can be neglected, allows for the sustainability of the CPRs: the consumers self-organize to reduce the extraction rate when the resource is low, and this is a sufficient mechanism to prevent the tragedy of the commons. If stochastic fluctuations on the resource dynamics are considered, then we might have CPRs collapse.

On the other hand, social interactions among players can generate rich and diverse patterns of social organization and polarization, which can have significant implications for the governance of CPR. 
In particular, both \textbf{QBS} and \textbf{EBS} can either enhance or hinder sustainability, depending on the connectivity and heterogeneity of the network. 
We have shown that there is a critical network mean degree at which a depletion/repletion phase transition occurs. 
This transition is accompanied by an absorbing/active state like of transition in the social dynamics.

The role of hubs in facilitating defection and depletion, or the effect of environmental feedback in promoting cooperation and repletion, are two main results of our work.

To study the robustness and generality of our results, we have investigated different types of game-theoretic dynamics to capture the complex and adaptive behaviors of consumers who exploit CPRs, whose dynamics are modeled through a logistic equation. 
We have also considered different network topologies and initial conditions. 

We have also measured the degree of polarization in the network, which depends on both the node and link attributes, and we have found that lower network connectivity and extraction coefficients lead to higher polarization. In this regard, we have observed that even if the majority of society cooperates, the human-environmental system can be set in depleted phases.

From a methodological point of view, we have compared the results obtained with Monte-Carlo simulations with the analytical and numerical ones from the rate equations. Although the two methodologies have given the same qualitative results, the nature of the sustainable-depleted transition was different in these two approaches. In this regard, more theoretical analyses and investigations are needed. 
Other limitations remain and may define the directions for future research. One limitation is that we have assumed that consumers are homogeneous in their extraction coefficients and knowledge feedback strategies, which may not be realistic in some cases. 
A possible extension is to introduce heterogeneity among consumers in these aspects and study how it affects the co-evolutionary dynamics and outcomes.
Another limitation is that we have neglected the role of governance systems or institutions in regulating CPRs, which are known to play an important role in enhancing or undermining sustainability. 
A possible extension could thus incorporate governance systems or institutions as another sub-system in our model, and study how they interact with consumers and resources. 
A third limitation is that we have used a simple logistic equation to model the resource dynamics, which may not capture some important features of CPRs such as non-linearity, threshold effects, or resilience. A possible extension of our model is thus developing more realistic models of resource dynamics and study how they affect the co-evolutionary dynamics and outcomes.


\section*{Acknowledgments}
M.S. acknowledges the fellowship from the Department of Physics And Astronomy ‘G. Galilei’, University of Padova. 

\section*{Code availability}
The GitHub link which include some part of the repository of the codes that we used for reproducing the numerical results: \url{https://github.com/MeghdadSaeedian/human-environmental-system}


\section*{Appendices}
\counterwithin{figure}{section}
\counterwithin{equation}{section}
\appendix
\section{Interpretation of positive and negative social interactions}\label{Appendix_neg_pos}
Here, we provide the interpretations and respective roles for each of the six possible pair-connections $\{a,b,c,d,e,f\}$ in the update rule. The pair-connections $\{a,c,e\}$ represent positive interactions (socially satisfying), while $\{b,d,f\}$ represent negative interactions (socially unsatisfying).

\begin{itemize}
	\item Pair-connection $a$: Two cooperators are connected with negative link, such as intense rivalry, or distrust. 
	We regard this configuration as a unsatisfying pair-connection because they share a common noble goal (self-organization to achieve sustainable HES) and a negative link would hinder this goal. 
	Therefore, if they act rationally, this configuration would eventually change. Either, for the sake of a sustainable HES, the negative link turns into a positive one ($a$ to $b$), or, one of the individuals succumbs to temptation and switches its strategy from cooperation to defection ($a$ to $f$). 
	According to the update rules, the former transition is more likely to occur in resource scarcity conditions and the latter in resource abundance.
	
	\item Pair-connection $c$: Two defectors are connected with negative link. 
	The rationale for classifying this configuration as a unsatisfying pair-connection is related to the concept of knowledge feedback. 
	In resource abundance, negative link is pointless, whereas a positive one ($c$ to $d$) can make resource exploitation more efficient, so it is rational to interact positively. 
	However, in the case of resource scarcity, according to knowledge feedback, one of the individuals may decide to change its strategy from defection to cooperation ($c$ to $f$).
	
	\item Pair-connection $e$: Two consumers with different strategies are connected with positive link, such as collaboration or friendship. 
	This is a conflicting situation because a key condition for any collaboration is to have the same strategy and approach, otherwise they cannot work together coherently. 
	In this case, there are three ways to achieve a positive pair-connection. 
	According to knowledge feedback, in resource abundance people are more likely to become greedy ($e$ to $d$), and in resource scarcity, they tend to be more committed to self-organization. 
	The latter effort can be realized through the complementary transitions ($e$ to $f$) and ($e$ to $b$) equally. 
	In simple terms, ($e$ to $f$) can be interpreted as a situation in which the cooperator boycotts the defector. On the other hand, ($e$ to $b$) can be interpreted as a situation in which the cooperator convinces the defector to become a cooperator.
	
	\item Pair-connection $b$: Two cooperators are connected with positive link. Having a positive bond between two players with the same strategies prevents any tendency to change the node's or link's state.
	
	\item Pair-connection $d$: Two defectors are connected with positive link. 
	Having a positive bond between two players with the same strategies prevents any tendency to change the node's or link's state.
	
	\item Pair-connection $f$: Two consumers with different strategies are connected with negative link, such as opposition or mutual boycott. 
	In fact, when two people have conflicting opinions or opposing strategies, the tendency is to view your opposition negatively rather than forge a positive connection built on understanding \cite{bliuc2015public}. 
\end{itemize}
All the above-mentioned statements are summarized in the update rule shown in Fig.~\ref{update_rules} (C) and Appendix.~\ref{Appendix_x_rho}. 

\section{Relationship between the density of cooperators x and the density of pair-connections $\pmb{\rho}$}\label{Appendix_x_rho}

Since the links among nodes in homogeneous random networks are generated randomly with probability $p=\frac{\mu}{N}$ (see equation (3.2) \cite{barrat2008dynamical}), we can relate the pair-connection densities, $\rho_i$, to the cooperator density $x(t)$.
For example, in an Erd\H{o}s-R\'enyi network with $N\gg1$, the $\rho_a+\rho_b$ is equal to the fraction of the number of pair-connections within the cooperators over the total number of pair-connections in the network. 
The former is $\frac{n(n-1)p}{2}$ and the latter is $\frac{N(N-1)p}{2}$, where $n$ is the number of cooperators. 
The density of pair-connections within the defectors, $\rho_c+\rho_d$, is obtained in a similar way using the number of defectors $(N-n)$.
The density of interface pair-connections between the cooperators and defectors, $\rho_e+\rho_f$, can be obtained by subtracting $\rho_a+\rho_b+\rho_c+\rho_d$ from one. 
All these relations are as follows \cite{saeedian2019absorbing}:
\begin{eqnarray}  \nonumber 
	\rho_{a}+\rho_{b}&\simeq&\frac{n(n-1)}{N(N-1)} \xrightarrow{n,N\gg1}x^2  \\ \nonumber 
	\rho_{c}+\rho_{d}&\simeq&\frac{(N-n)(N-n-1)}{N(N-1)}\xrightarrow{n,N\gg1}(1-x)^2 \\ \nonumber 
	\rho_{e}+\rho_{f}&\simeq&\frac{N(N-1)-(N-n)(N-n-1)-n(n-1)}{N(N-1)}\xrightarrow{n,N\gg1}2x(1-x) \\ \label{x_pair_rel} 
\end{eqnarray}
where $x=\frac{n}{N}$. Since the rate equations mimics the system in the thermodynamic limit, the approximate equality sign ($\simeq$) becomes an exact equality ($=$). Therefore, the final relation can be written as follows
\begin{eqnarray}
	x&=&\rho_{a}+\rho_{b}+\frac{\rho_{e}+\rho_{f}}{2}=1-\Big(\rho_{c}+\rho_{d}+\frac{\rho_{e}+\rho_{f}}{2}\Big). 
\end{eqnarray}

\section{Master equation and rate equations}\label{Appendix_ME_RE}

\begin{figure}[ht]
	\centering
	\includegraphics[width=.8\textwidth]{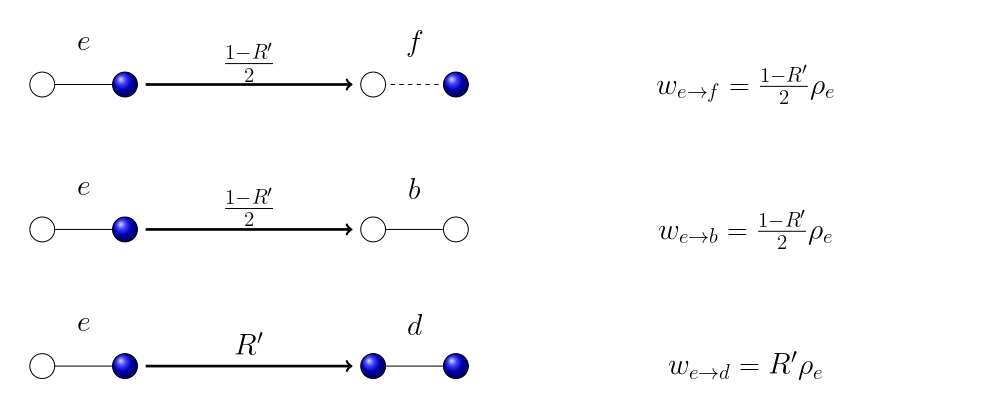}
	\caption{The conversion rates of the pair-connections according to the update rule applied in the Monte-Carlo  simulation. For illustration, we show $w_{e\rightarrow \omega}$ where $\omega \in (b,d,f)$. We explain the first transition ($w_{e\rightarrow f}$) in detail. Given that a pair-connection $e$ with probability $\rho_e$ is randomly chosen for updating, it changes to pair-connection $f$ with probability $\frac{1-R^{\prime}}{2}$. Thus, the transition probability is $\mathpzc{p}=\frac{1-R^{\prime}}{2}\rho_e$. Furthermore, since this conversion affects only one pair-connection, the event frequency per unit time is $\mathpzc{f}=1$. Hence, $w_{e\rightarrow f}=\frac{1-R^{\prime}}{2}\rho_e$. The top panel indicate a link-update, while the middle and bottom panels capture the node-updates.}
	\label{e_to_i_direct}
\end{figure}

\begin{figure}[ht]
	\centering
	\includegraphics[width=.8\textwidth]{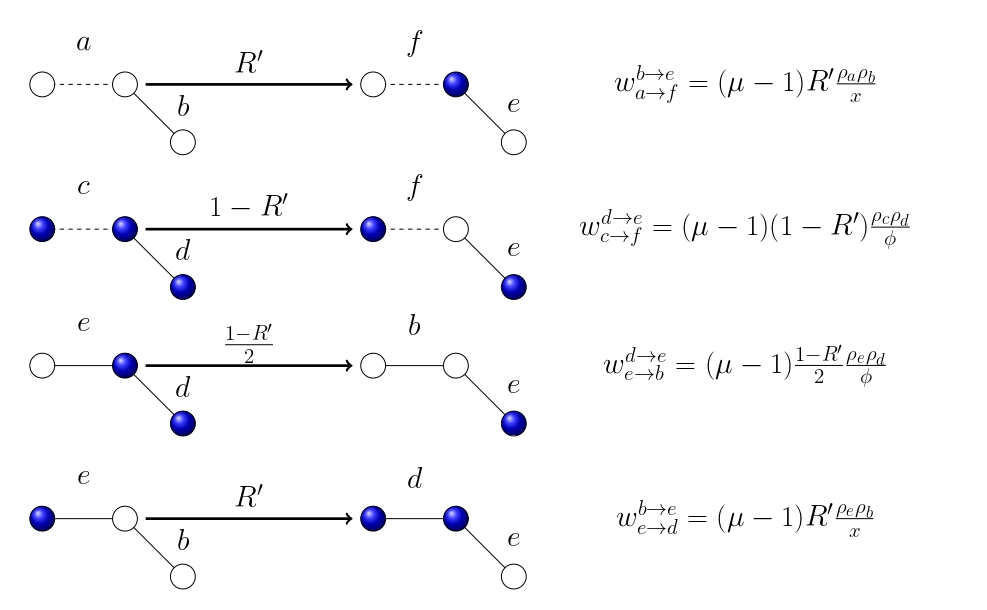}
	\caption{The conversion rates of the pair-connections according to the update rule induced in the Monte-Carlo simulation. For illustration, we show $w_{\nu \rightarrow \omega}^{i\rightarrow e}$ where $i \in (a,b,c,d,e,f)$. We explain the first transition ( $w_{a \rightarrow f}^{b\rightarrow e}$ ) in detail. Given that a pair-connection $a$ with probability $\rho_a$ is randomly chosen for updating, it changes to pair-connection $f$ with probability $R^{\prime}$. However, the pair-connection $b$ can be linked to the pair $a$ through the updated node. In the homogeneous network, the probability that a pair-connection $b$ is connected to a white node (cooperator) is $\frac{\rho_b}{x}$, as shown in Eq.~\ref{x_pair}. This is because, the only pair-connections that can be linked to a white node (cooperator) are $\{a,b,e,f\}$, but the contributions of these connections for $e$ and $f$ are half of $a$ and $b$. Therefore, $\mathpzc{p}=R^{\prime}\rho_a\frac{\rho_b}{x}$. Furthermore, since each node has a $\mu$ neighbor, node-update on a pair-connection (e.g. Fig.~(\ref{e_to_i_direct}) middle and bottom) will indirectly affect all other $\mu-1$ pair-connections linked to the updated node. Thus, the frequency of occurrence of this event per unit of time is $\mathpzc{f}=\mu-1$. Finally, $w_{a \rightarrow f}^{b\rightarrow e}=(\mu-1)R^{\prime}\frac{\rho_a \rho_b}{x}$.}
	\label{i_to_e}
\end{figure}

\begin{figure}[ht]
	\centering
	\includegraphics[width=.8\textwidth]{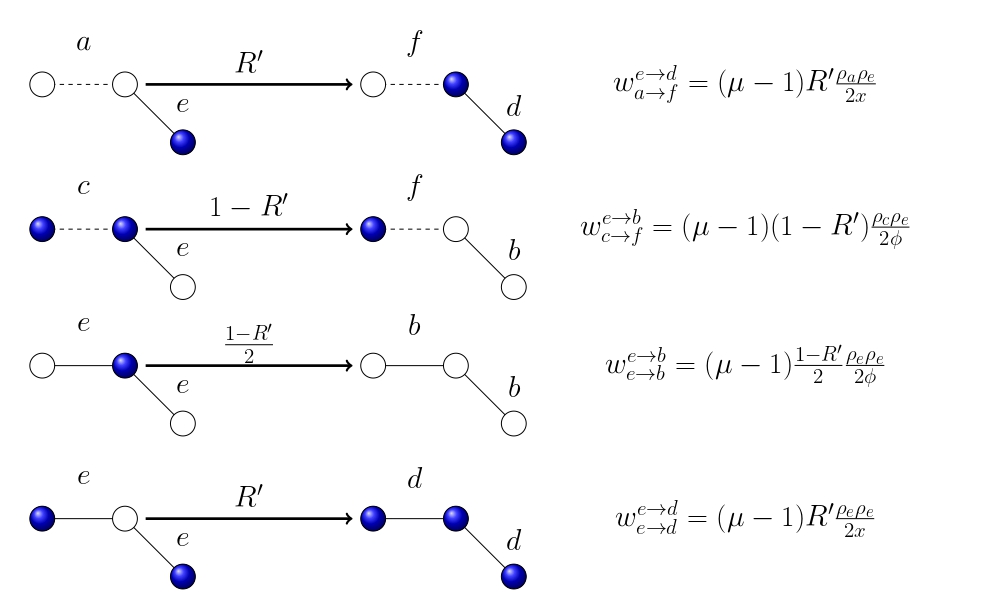}
	\caption{The conversion rates of the pair connections according to the update rule induced in the Monte-Carlo simulation. For illustration, we show $w_{\nu \rightarrow \omega}^{e\rightarrow i}$ where $i \in (a,b,c,d,e,f)$. We explain the second transition from top ( $w_{c \rightarrow f}^{e\rightarrow b}$ ) in detail. Given that a pair-connection $c$ with probability $\rho_c$ is randomly chosen for updating, it changes to pair-connection $f$ with probability $1-R^{\prime}$. However, the pair-connection $e$ is linked to the pair $c$ through a blue node. In the homogeneous network, the probability that a pair-connection $e$ is connected to a blue node (defector) is $\frac{\rho_e}{2\phi}$, as shown in Eq.~\ref{x_pair}. This is because, the only pair-connections that can be linked to a blue node (defector) are $\{c,d,e,f\}$, but the contributions of these connections of type $e$ and type $f$ are half of the $d$ and $c$ types. Therefore, $\mathpzc{p}=(1-R^{\prime})\rho_c\frac{\rho_e}{2\phi}$. Furthermore, since each node has a $\mu$ neighbor, node-update on a pair-connection (e.g. Fig.~(\ref{e_to_i_direct}) middle and bottom) will indirectly affect all other $\mu-1$ pair-connections linked to the updated node. Thus, the event frequency per unit time is $\mathpzc{f}=\mu-1$. Finally, $w_{c \rightarrow f}^{e\rightarrow b}=(\mu-1)(1-R^{\prime})\frac{\rho_c \rho_e}{2\phi}$.}
	\label{e_to_i}
\end{figure}

We can use formalisms like the master equation to describe the internal fluctuations, which are generated by the dynamics itself as a result of their interactions and are thus intrinsic to the dynamics \cite{horsthemke1977non}. 
This description suits the nature of the evolution of the pair-connections shown in Fig.~\ref{update_rules}. 
Therefore, we attempt to write down a master equation for the time-evolution of pair-connections. 
This dynamics is equivalent to a hypothetical multi-variable birth-death Markov process. 
For example, in the Fig.~\ref{update_rules} (C) for the Monte-Carlo process, when a pair-connection $e$ is selected for updating, technically, one pair-connection $e$ vanishes (disappears) and simultaneously either of pair-connections $b$, $d$, or $f$ with associated probabilities emerges.
This conversion occurs through two types of updating: link-update and node-update. 
Link updates only affect one pair-connection at a time (e.g. Fig.~\ref{e_to_i_direct}). 
However, in the node-updates, the status of all pair-connections connected to the updated node are affected (e.g. Figs.~\ref{i_to_e} and \ref{e_to_i}). 
When a node-update occurs, one pair-connection is converted due to the direct execution of the update-rule. 
In addition, $\mu-1$ pair-connections connected to the updated node are converted due to the indirect execution of the update-rule (e.g. Figs.~\ref{i_to_e} and \ref{e_to_i}). 
For the mean-field modeling of the latter case, at any moment in the dynamics, we assume that the pair-connections on the network are uniformly distributed. 
We observed that this approximation works well.
In this appendix, we provide details on how to write down the rate equations for the update rule shown in Fig.~\ref{update_rules}(C). 
A similar approach can be used for the update rule Fig.~\ref{update_rules}(B).

The first step to write down the master equation is to define the system: the system of interest here is the vector of the number of pair-connections on homogeneous networks. 
In homogeneous random networks and large Erd\H{o}s-R\'enyi networks (thermodynamic limit), the total number of links is $L=N\mu/2$ ($\mu$ is the mean degree of the network) such that
\begin{eqnarray}
	L&=&L_a+L_b+L_c+L_d+L_e+L_f
\end{eqnarray}
where $L_i, \quad i \in \{a,b,c,d,e,f\}$ is the number of pair-connections. Therefore, we can define a multi-variable system using the $L_i$'s as follow
\begin{eqnarray}
	\mathbf{L}&=&\{L_a,L_b,L_c,L_d,L_e,L_f\}, 
	\label{Multi_var_sys}
\end{eqnarray}
and, a single jumps out (in) of (to) the system $\mathbf{L}$, through the variable $L_i$ can be define by $\mathbf{L}-\mathbf{r}^{\pm}_i$, where
\begin{eqnarray}
	{\mathbf{r}}_i^{+}&=&+\{\delta_{a,i},\delta_{b,i},\delta_{c,i},\delta_{d,i},\delta_{e,i},\delta_{f,i}\}\cr
	{\mathbf{r}}_i^{-}&=&-\{\delta_{a,i},\delta_{b,i},\delta_{c,i},\delta_{d,i},\delta_{e,i},\delta_{f,i}\}\quad i \in \{a,b,c,d,e,f\}
\end{eqnarray}

\begin{eqnarray}
	\delta_{j,i} =
	\begin{cases}
		1 & i=j\\
		0 & \text{otherwise.}
	\end{cases}       
\end{eqnarray}
The ${\mathbf{r}}_i^{
	+}$'s are the unit basis vectors. The $\mathbf{L}$ and ${\mathbf{r}}_i^{\pm}$ are extensive quantities. The corresponding intensive quantities are defined here by $\pmb{\rho}$ and $\pmb{\epsilon}^{\pm}_i$, where 
\begin{eqnarray}
	\pmb{\rho}&=&\{\rho_a,\rho_a,\rho_a,\rho_c,\rho_d,\rho_e,\rho_f\}, \quad \rho_i=\frac{L_i}{L}
\end{eqnarray}
and
\begin{eqnarray}
	\pmb{\epsilon}_i^{+}&=&+\epsilon\{\delta_{a,i},\delta_{b,i},\delta_{c,i},\delta_{d,i},\delta_{e,i},\delta_{f,i}\}\cr
	\pmb{\epsilon}_i^{-}&=&-\epsilon\{\delta_{a,i},\delta_{b,i},\delta_{c,i},\delta_{d,i},\delta_{e,i},\delta_{f,i}\}\quad i \in \{a,b,c,d,e,f\}
\end{eqnarray}
where $\epsilon=\frac{1}{L}$ and note $\pmb{\epsilon}_{i}^{\pm}$'s are along the unit base vectors.

Therefore, the time evolution of the probability to find the system at state $\mathbf{L}$ can be written as
\begin{eqnarray} \nonumber
	\frac{\partial P(\mathbf{L},t)}{\partial t} =&\sum_{i}\Bigg(\sum_{{\mathbf{r}}_i^{\pm}} W(\mathbf{L}-{\mathbf{r}}_i^{\pm},{\mathbf{r}}_i^{\pm},t)P(\mathbf{L}-{\mathbf{r}}_i^{\pm},t) \\ &-\sum_{{\mathbf{r}}_i^{\pm}} W(\mathbf{L},{\mathbf{r}}_i^{\pm},t)P(\mathbf{L},t)\Bigg)
	\label{ME_ex}
\end{eqnarray}
where $W(\mathbf{L}-{\mathbf{r}}_i^{\pm},{\mathbf{r}}_i^{\pm},t)$ is the transition probability per unit time to pass state $\mathbf{L}-{\mathbf{r}}_i^{\pm}$ to $\mathbf{L}$ (Consequently, $W(\mathbf{L},{\mathbf{r}}_i^{\pm},t)$ is to pass state $\mathbf{L}$ to $\mathbf{L}+{\mathbf{r}}_i^{\pm}$). 
As seen, the superscript $+$ ($-$) always indicates an increasing (a decreasing) event in density of pair connection $i$. Therefore, the rate $W(\dots)$ holding ${\mathbf{r}}$ with superscript + (-) implies an increasing (decreasing) rate. 
The transition probability per unit time is often expressed as a scaling law of the form below based on a dimensionless parameter $L$ \cite{hanggi1978derivations} (see also equation II.3 of ref. \cite{horsthemke1977non})
\begin{eqnarray}\nonumber
	W(\mathbf{L}-{\mathbf{r}}_i^{\pm},{\mathbf{r}}_i^{\pm},t)&=&f(L)L\Bigg[\psi_0\Big(\pmb{\rho}-\pmb{\epsilon}^{\pm}_i,\pmb{\epsilon}^{\pm}_i,t\Big)+\epsilon \psi_1\Big(\pmb{\rho}-\pmb{\epsilon}^{\pm}_i,\pmb{\epsilon}^{\pm}_i,t\Big)+ \dots \Bigg]
\end{eqnarray}
Lets $f(L)=1$, and consider the fact that $P(\rho,t)= P(\mathbf{L},t)$ (see equation II.4 of Ref. \cite{horsthemke1977non}). 
The master equation Eq.~\ref{ME_ex} for the intensive macro-variable $\pmb{\rho}$ can be written as
\begin{eqnarray}
	\epsilon\frac{\partial P(\pmb{\rho},t)}{\partial t}&=&\sum_{i}\Bigg(\sum_{\pmb{\epsilon}^{\pm}_i} \Big[\psi_0\Big(\pmb{\rho}-\pmb{\epsilon}^{\pm}_i,\pmb{\epsilon}^{\pm}_i,t\Big)+\epsilon\psi_1\Big(\pmb{\rho}-\pmb{\epsilon}^{\pm}_i,\pmb{\epsilon}^{\pm}_i,t\Big)+ \dots \Big]P(\pmb{\rho} -\pmb{\epsilon}^{\pm}_i,t) \cr  &&-\sum_{\pmb{\epsilon}^{\pm}_i} \Big[\psi_0\Big(\pmb{\rho},\pmb{\epsilon}^{\pm}_i,t\Big)+\epsilon\psi_1\Big(\pmb{\rho},\pmb{\epsilon}^{\pm}_i,t\Big)+ \dots \Big]P(\pmb{\rho},t)\Bigg)
	\label{ME_in}
\end{eqnarray}

Assuming that each of the function $\psi_0\Big(\pmb{\rho}-\pmb{\epsilon}^{\pm}_i,\pmb{\epsilon}^{\pm}_i,t\Big)$ has Taylor expansion with respect to the first argument (So far, we follow the derivation in section III of ref. \cite{hanggi1978derivations}). 
Hence we obtain the Kramers-Moyal expansion (Equation 4.18 of \cite{garcia2007introduction}) for multivariate system Eq.~\ref{ME_in} as
\begin{eqnarray}
	\frac{\partial P(\pmb{\rho},t)}{\partial t}&=&\sum_{m=1}^{\infty}\frac{(-1)^m}{m!}\epsilon^{m-1}\sum_{{j_1}\dots {j_m}}\frac{\partial}{\partial \rho_{j_1} \dots \partial \rho_{j_m}}A^{(m)}_{{j_1}\dots {j_m}}(\pmb{\rho},t)P(\pmb{\rho},t)
	\label{ME_KM}
\end{eqnarray}
where  $j_{\nu} \in \{a,b,c,d,e,f\}$ and $\nu=1 \dots m$
\begin{eqnarray}\nonumber
	A^{(m)}_{{j_1}\dots {j_m}}(\pmb{\rho},t)&=&\sum_i\Bigg(\sum_{\pmb{\epsilon}^{\pm}_i} (r_{j_1}\dots r_{j_m})\Big[\psi_0\Big(\pmb{\rho},\pmb{\epsilon}^{\pm}_i,t\Big)+\epsilon\psi_1\Big(\pmb{\rho},\pmb{\epsilon}^{\pm}_i,t\Big)+ \dots \Big]\Bigg)
\end{eqnarray}
where $r_{j_{\nu}}$ ($\nu=1 \dots m$) is the $j^{th}$ components of the vector ${\mathbf{r}}_i^{\pm}$ ($\pmb{\epsilon}^{\pm}_i=\epsilon{\mathbf{r}}_i^{\pm}$). Since ${\mathbf{r}}_i^{+}$'s are unit basis vectors, the multiplication $(r_{j_1}\dots r_{j_m})$ is always zero except the case that $j_{\nu}$'s are the same. This leads to
\begin{eqnarray}
	\sum_{{j_1}\dots {j_m}}\frac{\partial}{\partial \rho_{j_1} \dots \partial \rho_{j_m}}\sum_i\sum_{\pmb{\epsilon}^{\pm}_i} (r_{j_1}\dots r_{j_m}) \longrightarrow\sum_j\Big(\frac{\partial}{\partial \rho_{j}}\Big)^m\sum_{\pmb{\epsilon}^{\pm}_j} ({\pm1})^{m}.
\end{eqnarray}
Note, index $j$ appears in $\psi_0\Big(\pmb{\rho},\pmb{\epsilon}^{\pm}_i,t\Big)$ and its higher orders. Therefore, Eq.~\ref{ME_KM} is reduced to  
\begin{eqnarray}
	\frac{\partial P(\pmb{\rho},t)}{\partial t}&=&\sum_{m=1}^{\infty}\frac{(-1)^m}{m!}\epsilon^{m-1}\sum_{j}\Big(\frac{\partial}{\partial \rho_{j}}\Big)^mA^{(m)}_j(\pmb{\rho},t)P(\pmb{\rho},t)
	\label{KM_sim}
\end{eqnarray}
where  
\begin{eqnarray}
	A^{(m)}_j(\pmb{\rho},t)&=&\sum_{\pmb{\epsilon}^{\pm}_j} (\pm1)^m\Big[\psi_0\Big(\pmb{\rho},\pmb{\epsilon}^{\pm}_j,t\Big)+\epsilon\psi_1\Big(\pmb{\rho},\epsilon^{\pm}_j,t\Big)+ \dots \Big]
	\label{A_rho}
\end{eqnarray}
Therefore, truncating the terms for $m>2$, the Fokker-Plank equation for multi-variable system of $\pmb{\rho}$ can be obtained as follow
\begin{eqnarray}
	\frac{\partial P(\pmb{\rho},t)}{\partial t}&=&-\sum_{j}\frac{\partial}{\partial \rho_{j} }A^{(1)}_j(\pmb{\rho},t)P(\pmb{\rho},t)+  \cr
	&&\frac{\epsilon}{2}\sum_{j}\Big(\frac{\partial}{\partial \rho_{j} }\Big)^2A^{(2)}_j(\pmb{\rho},t)P(\pmb{\rho},t)
	\label{FPE}
\end{eqnarray}
where the drift vector, $A^{(1)}_j(\pmb{\rho},t)$, and the diagonal component of diffusion matrix, $A^{(2)}_j(\pmb{\rho},t)$, are as follow 
\begin{eqnarray}
	A^{(1)}_j(\pmb{\rho},t)&=&\psi_0\Big(\pmb{\rho},\pmb{\epsilon}^{+}_j,t\Big)-\psi_0\Big(\pmb{\rho},\pmb{\epsilon}^{-}_j,t\Big) \cr \nonumber
	A^{(2)}_j(\pmb{\rho},t)&=&\psi_0\Big(\pmb{\rho},\pmb{\epsilon}^{+}_j,t\Big)+\psi_0\Big(\pmb{\rho},\pmb{\epsilon}^{-}_j,t\Big)
	\label{drift_diffusion}
\end{eqnarray}
Since the diffusion matrix has only diagonal elements, we can write the Ito stochastic differential equation for Eq.~\ref{FPE} using the formulas in Gardiner \cite{gardiner1985handbook}, equations 4.3.41 and 4.3.42.
\begin{eqnarray}
	\frac{d\rho_{j}(t)}{dt}=A_{j}^{(1)}(\pmb{\rho},t)+\sqrt{\epsilon A_{j}^{(2)}(\pmb{\rho},t)}\zeta
	\label{langevin}
\end{eqnarray}
where $\zeta$ is a multi-variable Wiener process. As evident, for a large size system $\epsilon \rightarrow 0$, the second term in the right side is negligible. Therefore,  we arrive at a well defined rate equations as (see also equation 3.5.19 of Gardiner \cite{gardiner1985handbook})
\begin{eqnarray}
	\frac{d\rho_{j}(t)}{dt}=\psi_0\Big(\pmb{\rho},\pmb{\epsilon}^{+}_j,t\Big)-\psi_0\Big(\pmb{\rho},\pmb{\epsilon}^{-}_j,t\Big)
	\label{re_general}
\end{eqnarray}

The transition probability per unit time $\psi_0\Big(\pmb{\rho},\pmb{\epsilon}^{+(-)}_j,t\Big)$ is the total rate by which the density of variable $\rho_j$ is  increasing (decreasing). The total rate is calculated by adding up all the rates by which jumping events occur per unit of time, as follow
\begin{eqnarray}
	\psi_0\Big(\pmb{\rho},\pmb{\epsilon}^{+}_j,t\Big)=\sum_{\nu}w_{\nu \rightarrow j}+\sum_{\nu \omega i}w_{\nu \rightarrow \omega}^{i\rightarrow j},  
	\quad \nu \in \{a, c, e\}, \quad \omega \in \{b, d, f\}, \quad i \in \{a,b,c,d,e,f\}  \nonumber
\end{eqnarray}
\begin{eqnarray}
	\psi_0\Big(\pmb{\rho},\pmb{\epsilon}^{-}_j,t\Big)=\sum_{\omega}w_{j \rightarrow \omega}+\sum_{\nu \omega i}w_{\nu \rightarrow \omega}^{j\rightarrow i},  
	\quad \nu \in \{a, c, e\}, \quad \omega \in \{b, d, f\}.  \quad i \in \{a,b,c,d,e,f\}  \nonumber
\end{eqnarray}
where the $w_{\nu \rightarrow j}$ ($w_{j \rightarrow \omega}$) is the rate at which jumping events are made into (out of) the variable $\rho_j$ directly due to execution of the update rule from pair-connection $\nu$ to $j$ ($j$ to $\omega$). Examples of these rates can be seen in Fig.~\ref{e_to_i_direct}. 
The $w_{\nu \rightarrow \omega}^{i\rightarrow j}$ ($w_{\nu \rightarrow \omega}^{j\rightarrow i}$) is the rate at which jumping events are made into (out of) the variable $\rho_j$ indirectly due to the node-update from $\nu$ to $\omega$ where the pair-connection $i$ is connected $\nu$ and $j$ is connected $\omega$ ($j$ is connected $\nu$ and $i$ is connected $\omega$). 
Example of these transition probability rates can be seen in Figs.~\ref{i_to_e} and \ref{e_to_i}. 
Finally, by definition, the rates $w_{\nu \rightarrow \omega}^{i\rightarrow j}$ and $w_{i\rightarrow j}$ are obtained by multiplying the probability of the occurrence of jumping event, $\mathpzc{p}$, and the frequency of the occurrence of that event per unit of time, $\mathpzc{f}$. 
In other word $w=\mathpzc{f}.\mathpzc{p}$. Note, this approach relies on an approximation that performs well when correlations are not strong. Also, some of these rates are zero, for instance $w_{e \rightarrow d}^{b\rightarrow c}=0$, or $w_{c \rightarrow b}=0$.

For illustration, we calculate the rate equations Eq.~\ref{re_general} for the pair connection $e$ for which all the direct and indirect transition probability rate are shown in the Fig.~\ref{e_to_i_direct} and Figs.~\ref{i_to_e} and \ref{e_to_i}, respectively.
\begin{eqnarray}
	\frac{d\rho_{e}}{dt} &=&\psi_0\Big(\pmb{\rho},\pmb{\epsilon}^{+}_e,t\Big)-\psi_0\Big(\pmb{\rho},\pmb{\epsilon}^{-}_e,t\Big)  \nonumber \cr
	\frac{d\rho_{e}}{dt} &=&\Bigg[\Big(w^{b\rightarrow e}_{a\rightarrow f}+w^{d\rightarrow e}_{c\rightarrow f}+w^{d\rightarrow e}_{e\rightarrow b}+w^{b\rightarrow e}_{e\rightarrow d}\Big)\Bigg]-  \nonumber  \\
	&&\Bigg[\Big(w_{e\rightarrow f}+w_{e\rightarrow b}+w_{e\rightarrow d}\Big)+\Big(w^{e\rightarrow d}_{a\rightarrow f}+w^{e\rightarrow b}_{c\rightarrow f}+w^{e\rightarrow b}_{e\rightarrow b}+w^{e\rightarrow d}_{e\rightarrow d}\Big)\Bigg]  \nonumber
\end{eqnarray}

Finally, following the same approach for other pair-connections we arrive at
\begin{eqnarray}
\	\frac{d\rho_{a}}{dt}&=&-\rho_{a}+(\mu-1)\Bigg[R'\Big(-\frac{\rho_{a}^{2}}{x}-\frac{\rho_{e}\rho_{a}}{x}\Big)+(1-R')\Big(+\frac{\rho_{c}\rho_{f}}{2\phi}+\frac{1}{2}\frac{\rho_{e}\rho_{f}}{2\phi}\Big)\Bigg]  \cr
	\
	\frac{d\rho_{b}}{dt}&=& (1-R')\rho_{a}+\frac{(1-R')}{2}\rho_{e}+(\mu-1)\Bigg[R'\Big(-\frac{\rho_{a}\rho_{b}}{x}-\frac{\rho_{e}\rho_{b}}{x}\Big)+(1-R')\Big(\frac{\rho_{c}\rho_{e}}{2\phi}+\frac{1}{2}\frac{\rho_{e}^{2}}{2\phi}\Big)\Bigg] \cr
	\
	\frac{d\rho_{c}}{dt}&=&-\rho_{c}+(\mu-1)\Bigg[R'\Big(\frac{\rho_{a}\rho_{f}}{2x}+\frac{\rho_{e}\rho_{f}}{2x}\Big)+(1-R')\Big(-\frac{\rho_{c}^{2}}{\phi}-\frac{1}{2}\frac{\rho_{e}\rho_{c}}{\phi}\Big)\Bigg]  \cr
	\
	\frac{d\rho_{d}}{dt}&=& R'\rho_{c}+R'\rho_{e}+(\mu-1)\Bigg[R'\Big(\frac{\rho_{a}\rho_{e}}{2x}+\frac{\rho_{e}^{2}}{2x}\Big)+(1-R')\Big(-\frac{\rho_{c}\rho_{d}}{\phi}-\frac{1}{2}\frac{\rho_{e}\rho_{d}}{\phi}\Big)\Bigg] \cr
	\
	\frac{d\rho_{e}}{dt}&=& -\rho_{e}+(\mu-1)\Bigg[R'\Big(-\frac{\rho_{a}\rho_{e}}{2x}+\frac{\rho_{a}\rho_{b}}{x}-\frac{\rho_{e}^{2}}{2x}+\frac{\rho_{e}\rho_{b}}{x}\Big)  \nonumber  \\ &&+(1-R')\Big(\frac{\rho_{c}\rho_{d}}{\phi}-\frac{\rho_{c}\rho_{e}}{2\phi}+\frac{1}{2}\frac{\rho_{e}\rho_{d}}{\phi}-\frac{1}{2}\frac{\rho_{e}^{2}}{2\phi}\Big)\Bigg]  \cr
	\
	\frac{d\rho_{f}}{dt}&=&R'\rho_{a}+(1-R')(\rho_{c}+\frac{\rho_{e}}{2}) +(\mu-1)\Bigg[R'\Big(-\frac{\rho_{a}\rho_{f}}{2x}+\frac{\rho_{a}^{2}}{x}-\frac{\rho_{e}\rho_{f}}{2x}+\frac{\rho_{e}\rho_{a}}{x}\Big)  \nonumber  \\
	&&+(1-R')\Big(\frac{\rho_{c}^{2}}{\phi}-\frac{\rho_{c}\rho_{f}}{2\phi}+\frac{1}{2}\frac{\rho_{e}\rho_{c}}{\phi}-\frac{1}{2}\frac{\rho_{e}\rho_{f}}{2\phi} \Big)\Bigg] 
	\label{RE_extended}
\end{eqnarray}

For the quenched binary state dynamics (Fig.~\ref{update_rules}(B)), the rate equations can be derived as we did for Eq.~\ref{RE_extended}. Therefore, the rate equations for the quenched dynamics can be written as

\begin{eqnarray}
   \	\frac{d\rho_{a}}{dt}&=&-R'\rho_{a}+(\mu-1)\Bigg[-\frac{R'}{x}\Big(\rho_{a}+\rho_{e}\Big)\rho_{a}+\frac{(1-R')}{2\phi}\Big(\rho_{c}+\rho_{e}\Big)\rho_{f}\Bigg]  \cr
	\
	\frac{d\rho_{b}}{dt}&=& (1-R')\rho_{e}+(\mu-1)\Bigg[-\frac{R'}{x}\Big(\rho_{a}+\rho_{e}\Big)\rho_{b}+\frac{(1-R')}{2\phi}\Big(\rho_{c}+\rho_{e}\Big)\rho_{e}\Bigg] \cr
	\
	\frac{d\rho_{c}}{dt}&=&-(1-R')\rho_{c}+(\mu-1)\Bigg[\frac{R'}{2x}\Big(\rho_{a}+\rho_{e}\Big)\rho_{f}-\frac{(1-R')}{\phi}\Big(\rho_{c}+\rho_{e}\Big)\rho_{c}\Bigg]  \cr
	\
	\frac{d\rho_{d}}{dt}&=& R'\rho_{e}+(\mu-1)\Bigg[\frac{R'}{2x}\Big(\rho_{a}+\rho_{e}\Big)\rho_{e}-\frac{(1-R')}{\phi}\Big(\rho_{c}+\rho_{e}\Big)\rho_{d}\Bigg]\cr
	\
	\frac{d\rho_{e}}{dt}&=& -\rho_{e}+(\mu-1)\Bigg[\frac{R'}{x}\Big(\rho_{a}+\rho_{e}\Big)\Big(\rho_{b}-\frac{\rho_e}{2}\Big)+\frac{(1-R')}{\phi}\Big(\rho_{c}+\rho_{e}\Big)\Big(\rho_{d}-\frac{\rho_e}{2}\Big)\Bigg] \cr
	\
	\frac{d\rho_{f}}{dt}&=&R'\rho_{a}+(1-R')\rho_{c}+   \nonumber  \\ &&(\mu-1)\Bigg[\frac{R'}{x}\Big(\rho_{a}+\rho_{e}\Big)\Big(\rho_{a}-\frac{\rho_f}{2}\Big)+\frac{(1-R')}{\phi}\Big(\rho_{c}+\rho_{e}\Big)\Big(\rho_{c}-\frac{\rho_f}{2}\Big)\Bigg] 
	\label{RE_quenched}
\end{eqnarray}

\section{Analytical Solution for the Rate Equations}\label{Appendix_Analytic}

Analytical solutions are presented here for the coevolution of resource dynamics  
(Eq.~\ref{resource_dynamics}) and rate equations of \textbf{QBS} (Eq.~\ref{RE_quenched}) and \textbf{EBS} (Eq.~\ref{RE}). These solutions enable us to analytically study the type of transitions that occur. For \textbf{QBS}, we found an exact analytical solution for an important particular case of Eq.~\ref{RE_quenched} where all the quenched interactions are negative. This solution was then coupled in an exact way with the dynamics given by Eq.~\ref{resource_dynamics}.  For \textbf{EBS} we found an exact analytical solution for Eq.~\ref{RE}. This exact solution was then coupled  with the dynamics provided by Eq.~\ref{resource_dynamics} and solved numerically. 

\begin{figure}[ht]
	\centering
	\includegraphics[width=1\textwidth]{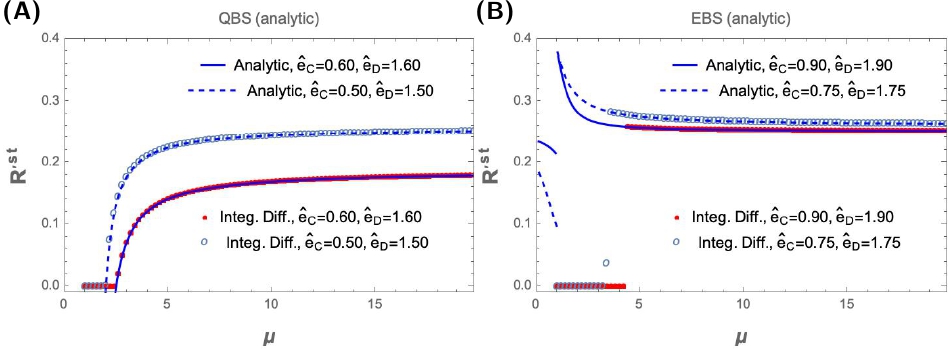}
	\caption{Comparison of analytical solution and numerical integration of differential equations for normalized resource levels as a function of mean degree. Panel (A) shows results for \textbf{QBS} (Eq.~\ref{R_mu_e_CD_q}) and panel (B) shows results for \textbf{EBS} (Eq.~\ref{R_mu_e_CD}). In both panels, there is excellent agreement between the analytical solution and numerical integration. The dynamics describing the system in panel (A) represent a particular case of general \textbf{QBS} dynamics which retains the key feature of a continuous phase transition, as shown in Fig.~\ref{results_R}(A). Thus panel (A) serves as an important approximation of the general case.  In Panel (B), the plateau amplitude derived analytically demonstrates quantitative agreement with the results from numerical integration, as evidenced by the black triangle marker in Fig.~\ref{results_R}(B). Finally, the discrepancy between the critical points obtained from the analytical and numerical integration methods is an area requiring further investigation. One potential explanation is that the non-zero analytical solutions (lines) in the regions that numerical integration (circles) yields zero ($R'^{st}=0$) is unstable.}
	\label{Analytic_Sus_1}
\end{figure}

In the case of \textbf{QBS}, the analytical results shown in Fig.~\ref{Analytic_Sus_1}(A) describe a particular system that retains the key property of a phase transition that is present in the general case. Thus, this particular case can be considered a close approximation of the general \textbf{QBS} dynamics, and can be studied in place of the more complex general system. The similarity is evident through comparison of Fig.~\ref{Analytic_Sus_1}(A) and Fig.~\ref{results_R}(A), which both exhibit the continuous phase transition characteristic of the general \textbf{QBS} case. In case \textbf{EBS}, the analytical results given by Fig.~\ref{Analytic_Sus_1}(B) are quantitatively in excellent agreement with the numerical integration of the differential equations presented by black triangle marker in Fig.~\ref{results_R}(B).

\subsection{A solution for co-evolution of recourse dynamics (Eq.~\ref{resource_dynamics}) and rate equation of \textbf{QBS} (Eq.~\ref{RE_quenched}) } \label{Appendix_Analytic_QBS}
We begin by expressing the derivative of the cooperator density, $\dot{x}$, with respect to the density of negative pair-connections.
Differentiating both sides of Eq.~\ref{x_pair} allows us to obtain the derivative of the cooperator density as:
\begin{eqnarray}
	\frac{dx}{dt}&=&\frac{d\rho_{a}}{dt}+\frac{d\rho_{b}}{dt}+\frac{1}{2}\Big(\frac{d\rho_{e}}{dt}+\frac{d\rho_{f}}{dt}\Big)\cr
	\
	\dot{x}&=&\mu\Big(-\frac{R'}{2}(\rho_{a}+\rho_{e})+\frac{1-R'}{2}(\rho_{c}+\rho_{e})\Big).
	\label{x_derivative_q}
\end{eqnarray}

Throughout this section, all variables are at their stationary states, however, for simplicity, we omit the superscript of stationary. 
Moreover, the case of $R=\rho_a=\rho_c=\rho_e=0$ is a trivial solution. In the following we focus on the solutions simultaneously positive $\rho_a+\rho_c+\rho_e>0$ and $R>0$ (replete and active phase).

At the stationary state, Eq.~\ref{x_derivative_q} gives a useful relation between the negative pair-connections as
\begin{eqnarray}
	\rho_c+\rho_e=\frac{R'}{1-R'}\Big(\rho_a+\rho_e\Big)
	\label{ce_ae_q}
\end{eqnarray}

In the second step, using  Eq.~\ref{ce_ae_q}, we try to find independent solution for $\rho_i$'s and $x$ in terms of $\mu$ and $R$.  The system of equations involving in this step can be sorted as

\begin{eqnarray}
\	0&=&-R'\rho_{a}+(\mu-1)R'(\rho_{a}+\rho_{e})\Bigg[-\frac{\rho_{a}}{x}+\frac{\rho_{f}}{2\phi}\Bigg]  \cr
	\
	0&=& (1-R')\rho_{e}+(\mu-1)R'(\rho_{a}+\rho_{e})\Bigg[-\frac{\rho_{b}}{x}+\frac{\rho_{e}}{2\phi}\Bigg] \cr
	\
	0&=&-(1-R')\rho_{c}+(\mu-1)R'(\rho_{a}+\rho_{e})\Bigg[\frac{\rho_{f}}{2x}-\frac{\rho_{c}}{\phi}\Bigg]  \cr
	\
	0&=& R'\rho_{e}+(\mu-1)R'(\rho_{a}+\rho_{e})\Bigg[\frac{\rho_{e}}{2x}-\frac{\rho_{d}}{\phi}\Bigg]\cr
	\
	0&=& -\rho_{e}+(\mu-1)R'(\rho_{a}+\rho_{e})\Bigg[\frac{1}{x}(\rho_{b}-\frac{\rho_{e}}{2})+\frac{1}{\phi}(\rho_{d}-\frac{\rho_{e}}{2})\Bigg] \cr
	\
	0&=&R'\rho_{a}+(1-R')\rho_{c}+(\mu-1)R'(\rho_{a}+\rho_{e})\Bigg[\frac{1}{x}(\rho_{a}-\frac{\rho_{f}}{2})+\frac{1}{\phi}(\rho_{c}-\frac{\rho_{e}}{2})\Bigg]  \cr
	\
	0&=&\mu\Bigg[-\frac{R'}{2}(\rho_{a}+\rho_{e})+\frac{1-R'}{2}(\rho_{c}+\rho_{e})\Bigg] \cr
	\
	0&=&-1+\rho_{a}+\rho_{b}+\rho_{c}+\rho_{d}+\rho_{e}+\rho_{f}
	\label{RE_set_inv_q}
\end{eqnarray}

While the analytical solution to Eq.\ref{RE_set_inv_q} is cumbersome, we can study a simplified scenario that still captures key system behaviors. Particularly, if we assume no positive connections exist ($\rho_b=\rho_e=\rho_d=0$), the system remains active while preserving symmetries in interaction types between the within-cooperators and within-defectors interactions. The analytical solution of this important particular case yields results very similar to the general case of \textbf{QBS} demonstrated in Fig.\ref{results_R}(A). Focusing on this particular case allows analytical tractability while eliciting the core system dynamics.

Substituting $\rho_e=0$ in Eq.\ref{RE_set_inv_q} and employing a standard software of mathematical analysis (Wolfram Mathematica, see section Code Availability), an independent analytical solution for $x$ is obtained as 
\begin{eqnarray}
x(R',\mu)=\frac{g_0(R',\mu)+\sqrt{g_1(R',\mu)}}{g_2(R',\mu)}
\label{x_R_mu_q}
\end{eqnarray}
where
\begin{eqnarray}
\ g_0(R',\mu)&=& -1 - 2 q + 2 {R'} + 2 q {R'}\cr \
g_1(R',\mu)&=&1 - 4 R + 4 q^2 {R'} + 4 {R'}^2 - 4 q^2 {R'}^2 \cr \
g_2(R',\mu)&=&2 (-1 - q + 2 {R'} + 2 q {R'})
 \label{g_i_q}
\end{eqnarray}
and q=$\mu-1$. The Eq.~\ref{x_R_mu_q} is a analytical solution for the rate equations Eq.~\ref{RE_quenched}. In order to find a solution for co-evolution of \textbf{QBS} (Eq.~\ref{RE_quenched}) and resource dynamics (Eq.~\ref{resource_dynamics})  we can substitute Eq.~\ref{x_R_mu_q} in the resource dynamics at the stationary state when $R>0$ as follow 
\begin{eqnarray}
0=1- (1-\hat{e}_C)R' - x(R',\mu) (\hat{e}_C-\hat{e}_D) -\hat{e}_D.
\label{R_mu_e_CD_q}
\end{eqnarray}

Solving Eq.\ref{R_mu_e_CD_q} analytically derives an independent relationship between $R'$ and the control parameters ($\mu$, $\hat{e}_C$, $\hat{e}_D$). Since this analytical solution is lengthy, it is presented in the GitHub repository Mathematica file detailing the full solution to Eq.\ref{R_mu_e_CD_q}. This separate file allows the interested reader to examine the full analytical details, while avoiding cluttering the current discussion. The key insight is that an independent functional connection between $R'$ and ($\mu$, $\hat{e}_C$, $\hat{e}_D$) emerges from analytically solving the dynamics of the co-evolving quenched system.

The solution to Eq.\ref{R_mu_e_CD_q} is plotted as a function of $\mu$ in Fig.\ref{Analytic_Sus_1}(A). There is evident qualitative agreement between the analytical solution of this particular case and the general continuous phase transition seen in Fig.~\ref{results_R}(A).

\subsection{A solution for co-evolution resource dynamics (Eq.~\ref{resource_dynamics}) and rate equation of \textbf{EBS} (Eq.~\ref{RE})}\label{Appendix_Analytic_EBS}
We first derive an exact analytical solution to Eq.\ref{RE}. This establishes the baseline social evolutionary dynamics in isolation. We then numerically solve for the co-evolution of Eq.\ref{RE} coupled to the resource dynamics captured by Eq.~\ref{resource_dynamics}. 

We start by expressing the derivative of the cooperator density in terms of the negative pair connections as
\begin{eqnarray}
	\frac{dx}{dt}&=&\frac{d\rho_{a}}{dt}+\frac{d\rho_{b}}{dt}+\frac{1}{2}\Big(\frac{d\rho_{e}}{dt}+\frac{d\rho_{f}}{dt}\Big)\cr
	\
	\dot{x}&=&\mu\Big(-\frac{R'}{2}(\rho_{a}+\rho_{e})+\frac{1-R'}{2}(\rho_{c}+\frac{\rho_{e}}{2})\Big).
	\label{x_derivative}
\end{eqnarray}
Next, we rewrite Eq.~\ref{x_derivative} at the stationary state and obtain a useful relation between the negative pair-connections as
\begin{eqnarray}
	\rho_c+\frac{\rho_e}{2}=\frac{R'}{1-R'}\Big(\rho_a+\rho_e\Big)
	\label{ce_ae}
\end{eqnarray}
In the second step, using  Eq.~\ref{ce_ae}, we try to find independent solution for $\rho_i$'s and $x$ in terms of $\mu$ and $R$. The system of equations required to find the independent solutions are as follow  
\begin{eqnarray}
\	0&=&-\rho_{a}+(\mu-1)R'(\rho_{a}+\rho_{e})\Bigg[-\frac{\rho_{a}}{x}+\frac{\rho_{f}}{2\phi}\Bigg]  \cr
	\
	0&=& (1-R')\rho_{a}+\frac{(1-R')}{2}\rho_{e}+(\mu-1)R'(\rho_{a}+\rho_{e})\Bigg[-\frac{\rho_{b}}{x}+\frac{\rho_{e}}{2\phi}\Bigg] \cr
	\
	0&=&-\rho_{c}+(\mu-1)R'(\rho_{a}+\rho_{e})\Bigg[\frac{\rho_{f}}{2x}-\frac{\rho_{c}}{\phi}\Bigg]  \cr
	\
	0&=& R'\rho_{c}+R'\rho_{e}+(\mu-1)R'(\rho_{a}+\rho_{e})\Bigg[\frac{\rho_{e}}{2x}-\frac{\rho_{d}}{\phi}\Bigg]\cr
	\
	0&=& -\rho_{e}+(\mu-1)R'(\rho_{a}+\rho_{e})\Bigg[\frac{1}{x}(\rho_{b}-\frac{\rho_{e}}{2})+\frac{1}{\phi}(\rho_{d}-\frac{\rho_{e}}{2})\Bigg] \cr
	\
	0&=&R'\rho_{a}+(1-R')(\rho_{c}+\frac{\rho_{e}}{2})+(\mu-1)R'(\rho_{a}+\rho_{e})\Bigg[\frac{1}{x}(\rho_{a}-\frac{\rho_{f}}{2})+\frac{1}{\phi}(\rho_{c}-\frac{\rho_{f}}{2})\Bigg]  \cr
	\
	0&=&\mu\Bigg[-\frac{R'}{2}(\rho_{a}+\rho_{e})+\frac{1-R'}{2}(\rho_{c}+\frac{\rho_{e}}{2})\Bigg] \cr
	\
	0&=&-1+\rho_{a}+\rho_{b}+\rho_{c}+\rho_{d}+\rho_{e}+\rho_{f}
	\label{RE_set_inv}
\end{eqnarray}
Employing a standard software of mathematical analysis (Wolfram Mathematica, see section Code Availability), an independent analytical solution for $x$ is obtained as 
\begin{eqnarray}
x(R',\mu)=\frac{g_0(R',\mu)-\sqrt{g_1(R',\mu)}}{g_2(R',\mu)}
\label{x_R_mu}
\end{eqnarray}
where
\begin{eqnarray}
\ g_0(R',\mu)&=& {R'} + 3 q R' - 7 R^2 - 8 q {R'}^2 + 13 {R'}^3 + q {R'}^3 - 4 {R'}^4 + 8 q {R'}^4 - 
 4 q R^5\cr
	\
g_1(R',\mu)&=&2 - 22 {R'} + 97 {R'}^2 - 9 q^2 {R'}^2 - 214 {R'}^3 + 84 q^2{R'}^3 + 235 {R'}^4 - 
 280 q^2 {R'}^4 \cr \ && - 98 {R'}^5 + 398 q^2 {R'}^5 - 17 {R'}^6 - 131 q^2 {R'}^6 + 22 {R'}^7 - 
 298 q^2 {R'}^7 - 4 {R'}^8  \cr \ && +388 q^2 {R'}^8 - 184 q^2 {R'}^9 + 32 q^2 {R'}^{10} \cr
	\
g_2(R',\mu)&=&-2 + 14 {R'} + 6 q {R'} - 36 {R'}^2 - 28 q {R'}^2 + 38 {R'}^3 + 40 q {R'}^3 - 10 {R'}^4 \cr \ && - 
 22 q {R'}^4 + 4 q {R'}^5
 \label{g_i}
\end{eqnarray}
and q=$\mu-1$. The Eq.~\ref{x_R_mu} is an exact analytical solution for the rate equations Eq.~\ref{RE}. 
In order to find a solution for co-evolution of \textbf{EBS} (Eq.~\ref{RE}) and resource dynamics (Eq.~\ref{resource_dynamics})  we can substitute Eq.~\ref{x_R_mu} in the resource dynamics at the stationary state when $R>0$ as follow 
\begin{eqnarray}
0=1- (1-\hat{e}_C)R' - x(R',\mu) (\hat{e}_C-\hat{e}_D) -\hat{e}_D.
\label{R_mu_e_CD}
\end{eqnarray}

Solving Eq.\ref{R_mu_e_CD} derives an independent relationship between $R'$ and the control parameters ($\mu$, $\hat{e}_C$, $\hat{e}_D$). The details to solving the Eq.\ref{R_mu_e_CD} presented in GitHub (see Code Availability). 
Note that by truncating on the lower orders in Eq.~\ref{g_i}, an approximate analytical solution can be obtained, but here we tried to keep all terms. 

The solution to Eq.~\ref{R_mu_e_CD}, shown in Fig.~\ref{Analytic_Sus_1}(B), displays discontinuity above the critical point. The solution also approaches a nonzero plateau whose amplitude quantitatively matches numerical integration (also see the black triangle marker in Fig.~\ref{results_R}(B)). However, further investigation is needed to clarify the discrepancy in critical points between the analytical and numerical integration approaches. 


\section{Fluctuation-driven absorbing states in finite size systems}\label{Appendix_tau}

\begin{figure}[h!]\centering
	\includegraphics[width=1\linewidth]{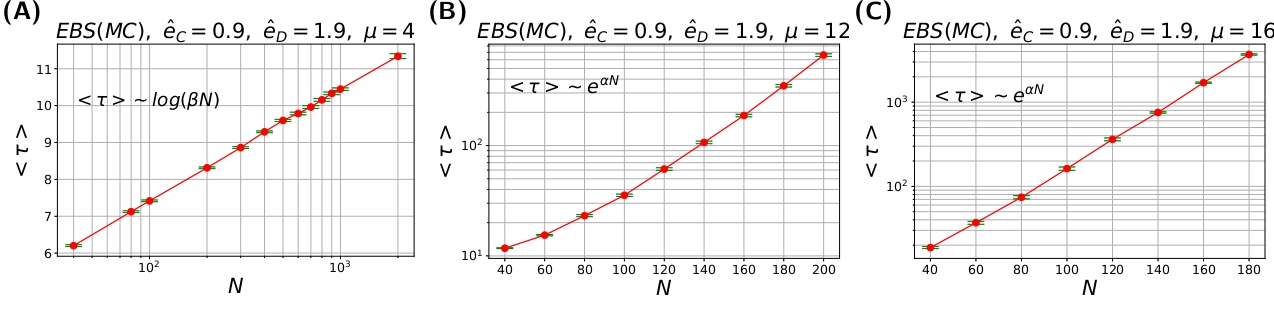}
	\caption{Average characteristic time to reach an absorbing state as a function of system size for two sets of parameters leading to either an absorbing phase in panel (A) or an active phase in panels (B) and (C). As shown, the curves in the absorbing (active) phase are approximately logarithmic (exponential). All the computed values are shown with their error bars.}
	\label{tau}
\end{figure}

The existence of absorbing states implies that a finite system will eventually fluctuate into absorbing state. However, the key is determining how the characteristic absorption time, $\tau$, depends on system size $N$. Fig.~\ref{tau} show the average of the characteristic time as a function of system size. The mean time $\mu$, plotted in panel A, representing relatively smaller mean degree ($<\mu>=4$), show a logarithmic increase with system size $<\tau>\sim \text{log}(\beta N)$, while in panels B and C ($<\tau>=12, 16$), grows exponentially with $N$ as $<\tau>\sim e^{\alpha N}$. For sufficiently large N, an exponential form reflects the rarity of absorption events. This exponential divergence signifies that decay to the absorbing state requires enormously long times for large systems. From a statistical mechanics perspective, the active phase persists indefinitely in the thermodynamic limit as $N\rightarrow \infty$. Fluctuation-driven absorption vanishes, and the active state is a true macroscopic phase, stabilized exponentially with size. 

\section{Effect of initial condition and network topology in the  depletion/repletion transition}\label{Appendix_IC}

\begin{figure}[h!]\centering
	\includegraphics[width=1\linewidth]{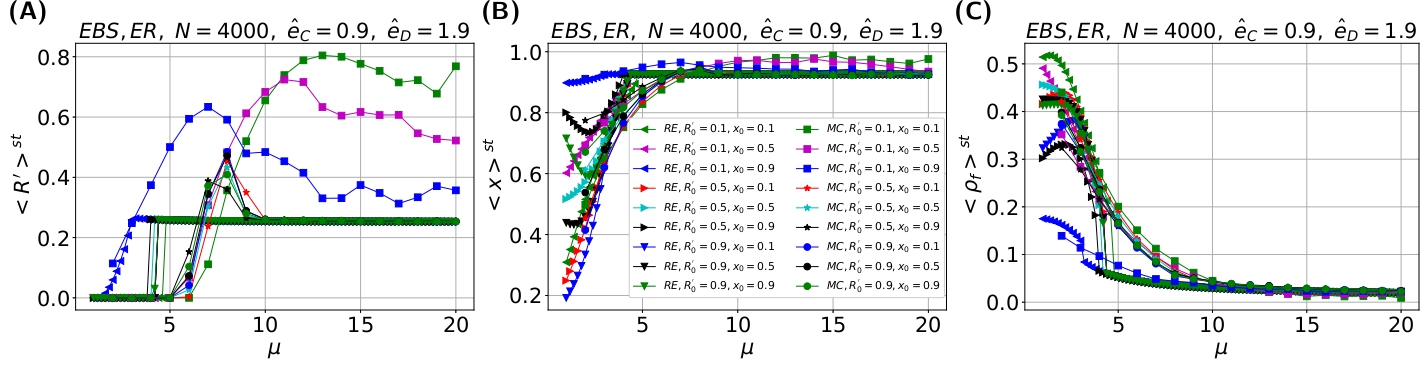}
	\caption{Effects of initial conditions on ${<R^{\prime}>}^{\text{st}}$, $<x>^{\text{st}}$, and $<\rho_f>^{\text{st}}$. Using the \textbf{ EBS} dynamics model outlined in Fig. \ref{update_rules}(C), we simulated across a broad spectrum of initial conditions on Erd\H{o}s-Rényi networks. Specifically, we systematically varied the initial settings of $x_0 \in \{0.1, 0.5, 0.9\}$ and $R'_0 \in \{0.1, 0.5, 0.9\}$, covering the nine combinations of these parameterized starting states. Based on these input configurations, we executed Monte-Carlo simulations alongside numerical integration of coupled differential rate equations governing the dynamics. The consistent legend is only presented in panel (B). In panel (A), we indicate the depletion/repletion transition in ${<R^{\prime}>}^{\text{st}}$. As is evident, in Monte Carlo simulation, all curves collapse in the same steady state, except those starting from $R^{\prime}_0=0.1$, where the trajectories fluctuate. Even these exceptional cases exhibit attraction to the common steady state statistics when examined under an expanded lens, by aggregating many stochastic realizations and simulating larger system sizes. In panel (B) and (C), we plot the ${<x>}^{\text{st}}$ and ${<\rho_f>}^{\text{st}}$, respectively. As evident, in the absorbing phase (here, $\mu \lessapprox 5$), the density of cooperators in (B) and the level of polarization in (C) depend on the initial condition $R'_0$ (for open access code see Code Availability section).}
	\label{results_R_initi_R}
\end{figure}

\begin{figure}[h!]\centering
	\includegraphics[width=1\linewidth]{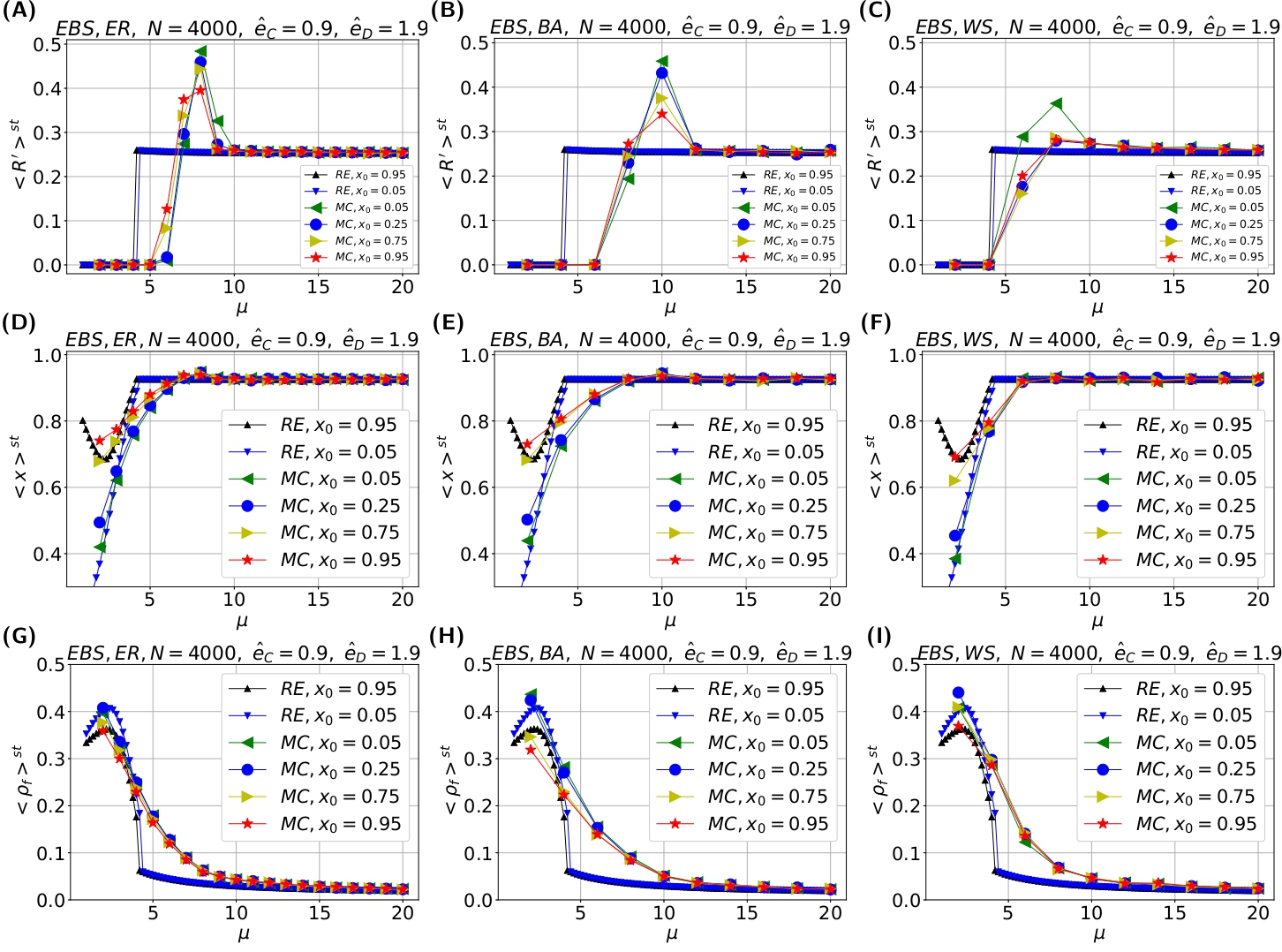}
	\caption{Effects of network topology on ${<R^{\prime}>}^{\text{st}}$ and $<x>^{\text{st}}$. (A) and (D) Results for the Erd\H{o}s-R\'enyi network. (B) and (E) Results for the Barabasi-Albert network. (C) and (F) Results for the Watts-Strogatz network. We show how these quantities vary with network mean degree $\mu$ for four initial cooperator densities ( $x_0 = 0.05, 0.25, 0.75, 0.95$ ) and three network topologies: Erd\H{o}s-R\'enyi, Barabasi-Albert and Watts-Strogatz (with rewiring probability 0.05). The initial values are $l_0=0.5$ and $R^{\prime}_0=\frac{2}{3}$ for all cases. The network topology does not qualitatively affect the deplete/replete transition, but there is a slight correlation between the critical values and the network heterogeneity. The Barabasi-Albert network, which is the most heterogeneous, has the largest critical value, while the Watts-Strogatz network, which is the least heterogeneous, has the smallest critical value. In the absorbing phase, ${<R^{\prime}>}^{\text{st}}$ and $<x>^{\text{st}}$ depend on the initial conditions, but in the active phase they do not. The Monte-Carlo and rate equations results agree well. (G), (H) and (I) ${<\rho_f>}^{\text{st}}$ as a function of $\mu$, which measures the level of polarization, $r=2\rho_f$. The ${<\rho_f>}^{\text{st}}$ depends on the initial conditions in the depleted phase, but not in the repleted phase. The depleted phase also shows a higher degree of polarization than the repleted phase.}
	\label{results_R_initi}
\end{figure}

The dynamic behavior of complex systems can depend sensitively on both the initial conditions and the structure of the underlying interaction network. Here we provide the phase transition diagram for the co-evolution \textbf{EBS} dynamics given in Fig.~\ref{update_rules} (C) for a wide range of initial condition in Fig.~\ref{results_R_initi_R} and three underlying network topology in Fig.~\ref{results_R_initi}.

\clearpage

\section{Depletion/Repletion transition in ($\hat{e}_C,\hat{e}_D$) phase space }\label{Appendix_e_D_e_C}
\begin{figure}[h!]\centering
	\includegraphics[width=1\linewidth]{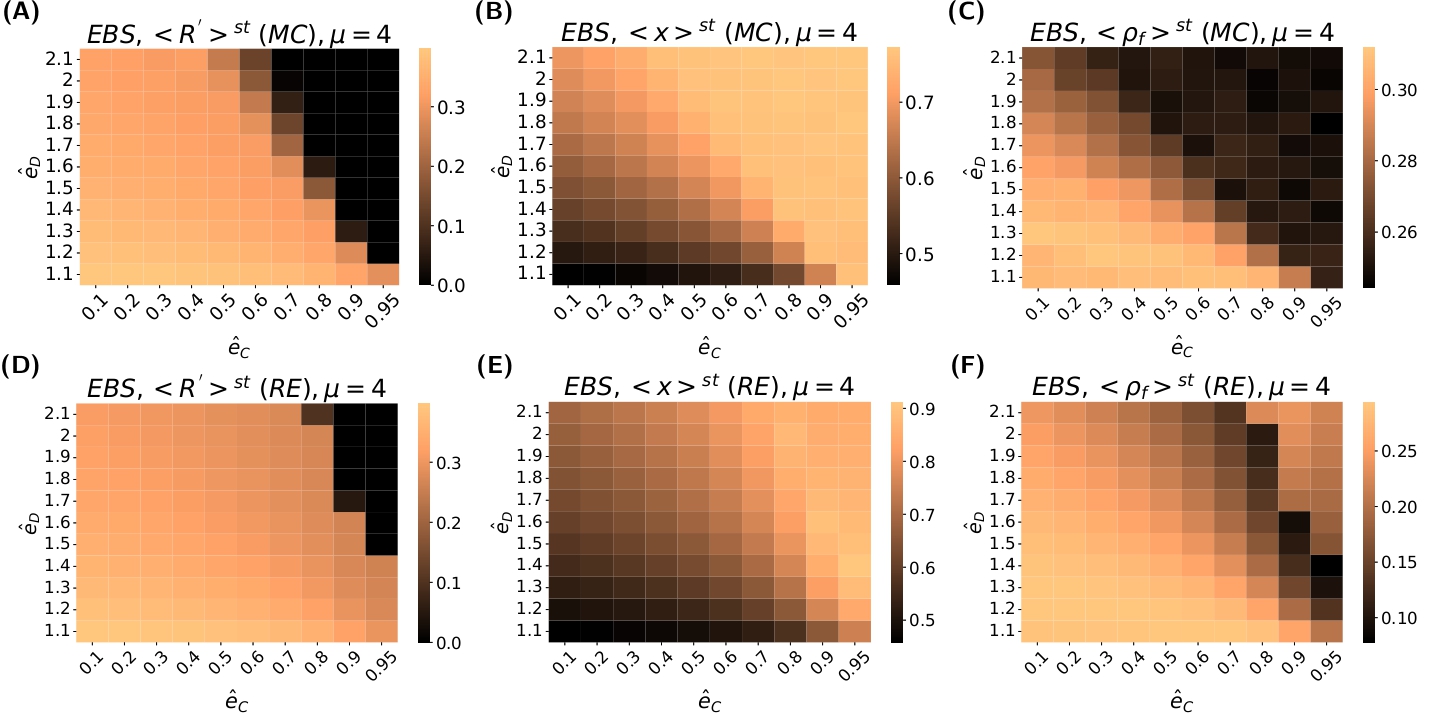}
	\caption{Depletion/Repletion transition in ($\hat{e}_C,\hat{e}_D$) phase space. In this figure, we plot the color-coded order parameters ( $R$, $x$, and $f$ ) as functions of the control parameters ( $\hat{e}_C$,$\hat{e}_D$, and $\mu=4$ ) in a raster plot. The simulations run for $N=4000$. The top (bottom) panels show the results from Monte Carlo (rate equations) simulations. (A) and (D) The deplete/replete transition in the plane of extraction coefficients ( $\hat{e}_C$ and $\hat{e}_D$ ). The deplete (replete) phase is identified where $<R'>^{st}=0$ ($<R'>^{st}>0$). In other words, a distinctive border between black and cooper regions marks the critical line of the transition. The panel (A) suggests a discontinuous phase transition due to the abrupt change in color codes at the transition boundary. Further simulations (not presented here) reveal the critical line shifts to larger values of $\hat{e}_C$ and $\hat{e}_D$ as $\mu$ increases. (B) and (E) It is almost always the case that the cooperators outnumber the population by more than half. This is scary because sometimes even above eighty percent of cooperators does not guarantee the sustainability HES. (C) and (F) Lower extraction coefficients lead to higher polarization.}
	\label{results_R_mu_fiex}
\end{figure}

The critical line of depletion/repletion transition is extended in phase space of ($\hat{e}_C,\hat{e}_D$) as presented in Fig.~\ref{results_R_mu_fiex}(A)\&(D). The essence of this transition seems to be discontinuous due to the abrupt change in the normalized level of resource at the stationary state, $<R'>^{st}$. In order to make an accurate claim on the type of transition, extensive finite size analysis is required. This is beyond the scope of this manuscript, so we leave it here for future research. 

In Fig.~\ref{results_R_mu_fiex}(B)\&(E), the average of density of cooperators is presented in color-code in ($\hat{e}_C,\hat{e}_D$) phase space. As evident, even in the depleted phase, the density of cooperators almost always builds the majority of the population ($x^{st}\gtrsim 0.5$). In other words, as also shown in the main text, even if the majority of the population cooporates, the HER system is not guaranteed to be sustainable. 

\begin{figure}[h!]\centering
	\includegraphics[width=1\linewidth]{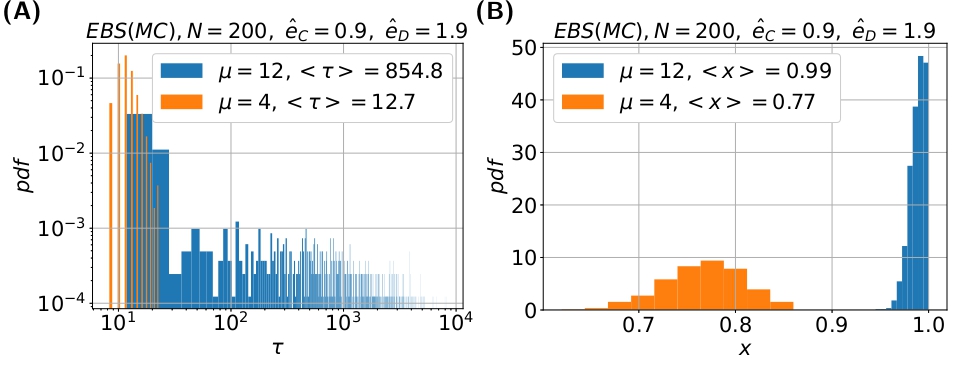}
	\caption{(A) shows the distribution of the characteristic absorption time $\tau$ for network mean degrees of $\mu=4$ and $\mu=12$, representing the absorbing and active phases respectively. (B) displays the distribution of the density of cooperators $x$ for the same two network conditions leading to the absorbing ($\mu=4$) and active ($\mu=12$) phases. Comparing the two panels illustrates how the absorptive dynamics that emerge at lower mean degrees impact both the time to reach a frozen state as well as the eventual composition of cooperators within the frozen states. Finally note, the distribution of the characteristic absorption time $\tau$ differs markedly between the active phase ($\mu=12$) and the absorbing phase. In the active phase, $\tau$ has an extended, multi-scaling distribution, unlike the more limited distribution seen in the absorbing phase. A log-log plot is therefore used in panel (A) to depict the $\tau$ distributions for both the active and absorbing phases on the same axes. This allows visualization of the full range of timescales exhibited in the active phase, which spans orders of magnitude longer than the absorption times typically observed in the absorbing phase.} 
	\label{tau_t_x}
\end{figure}

The case shown in Fig.~\ref{results_R_mu_fiex}(B) demonstrates a counterintuitive result - the density of cooperators in the depleted phase (characterized by high $\hat{e}_C$ and $\hat{e}_D$) is actually higher than the density of cooperators in the repleted phase. This seems paradoxical at first glance. However, if we plot the steady-state density of cooperators ($x^{st}$) as a function of the network mean degree ($\mu$) for any given combination of $\hat{e}_C$ and $\hat{e}_D$ values, we see that $x^{st}$ is uniformly lower in the absorbing phase than in the active phase. This trend is illustrated  in Fig. \ref{tau_t_x}(B). This illustrates that the density of cooperators decreases when transitioning from the active state to the absorbing state. However, the cooperator density remains higher than that of defectors in both phases. Consider the average values, for instance.

In Fig.~\ref{results_R_mu_fiex}(C)\&(F), we present the $<\rho_f>^{st}$ as index of polarization (see section 3.5). As seen, the lower extraction coefficients lead to higher polarization.

\section{Individual vs. social dynamics}\label{Appendix_in_so}
When the social network is densely connected, the social dynamics of the \textbf{QBS} \& \textbf{EBS} models depicted in Fig. \ref{update_rules}(B)\&(C) exhibit behavior similar to that of the individual dynamics \textbf{NS} shown in Fig. \ref{update_rules}(A).

By definition the \textbf{NS} is a dynamic solely on the nodes, neglecting the presence of interactions among users. This mechanism holds when a consumer makes a decision without the influence of other consumers. On the other hand, the case of \textbf{QBS} \& \textbf{EBS} is defined solely for the links that are conditioned by the strategy of consumers on their sides. These mechanisms are true when we attempt to model the dependence of a consumer's decision on the status of other consumers (link and node). 

The key concept of social interactions in \textbf{QBS} \& \textbf{EBS} is the reluctance of individual to update their strategy stemming from the positive bound with the like-minded friends (pair-connection $b$ and $d$) or negative connections with opponent (pair-connection $f$). Therefore, when there is a high density of $b$, $d$, and $f$ pair-connections within the network, the level of activity of both nodes and links tends to be low (and vice-versa). The update rules presented in Fig. \ref{update_rules}(B)\&(C) attempt to increase the density of positive pair-connections ($b$,$d$,$f$)  by converting the negative pair-connections ($a$,$c$,$e$) to the positive ones ($b$,$d$,$f$). However, the node-updates, as evidently shown in Fig.~\ref{i_to_e} and Fig.~\ref{e_to_i}, change the status of all pair-connections that are connected to the updated-nodes, slowing down the increase of positive interactions ($b$,$d$,$f$). This effect is even amplified when connectivity in the network increases. Therefore, increasing $\mu$ tends to stabilize the density of positive pair connections ($b$,$d$,$f$), by increasing the level of activity of the nodes in the social system. This makes the \textbf{QBS} \& \textbf{EBS}  similar to the dynamics of \textbf{NS} where there is no link to prevent nodes activity.  


\section{Initial condition}\label{Appendix_ic}

To ensure that the initial condition satisfies the relations, we use the following expression as the initial condition when integrating numerically
\begin{eqnarray}
	\rho_a(0)&=&x_0^2(1-\ell_0)\nonumber\\
	\rho_b(0)&=&x_0^2\ell_0\nonumber\\
	\rho_c(0)&=&(1-x_0)^2(1-\ell_0)\nonumber\\
	\rho_d(0)&=&(1-x_0)^2\ell_0 \nonumber\\
	\rho_e(0)&=&2x_0(1-x_0)\ell_0 \nonumber\\
	\rho_f(0)&=&2x_0(1-x_0)(1-\ell_0)
\end{eqnarray}
where $\ell_0$ is the fraction of friendly links.

\clearpage


\bibliographystyle{unsrt}
\bibliography{Draft.bib}
\end{document}